\documentclass[sigconf, nonacm, natbib=false]{acmart}

\renewcommand\footnotetextcopyrightpermission[1]{} 

\usepackage{booktabs}     
\usepackage{csquotes}
\usepackage{enumitem}     
\usepackage{array}        
\usepackage{tabularx}     
\usepackage{graphicx}
\usepackage{array}
\usepackage{booktabs}
\usepackage{algorithm}
\usepackage{algorithmic}
\usepackage{caption}
\usepackage{xcolor}
\usepackage{hyperref}
\hypersetup{
    colorlinks=true,
    linkcolor=blue,
    filecolor=magenta,      
    urlcolor=cyan,
    pdftitle={Overleaf Example},
    pdfpagemode=FullScreen,
    }
\usepackage{booktabs}  
\usepackage{tabularx}  

\usepackage[skins]{tcolorbox}
\usepackage{fontawesome5}

\newtcolorbox{prettiquote}{
    enhanced,
    boxrule=0pt,
    frame hidden,
    borderline west={2pt}{0pt}{gray!30},
    colback=gray!5,
    left=10pt,
    right=5pt,
    top=5pt,
    bottom=5pt,
    fontupper=\itshape,
    before upper={\textcolor{gray!50}{\Large\faQuoteLeft}\quad}
}

\AtBeginDocument{%
  }

\DeclareUnicodeCharacter{2217}{*}

\usepackage{algorithm} 
\usepackage{algorithmic}
\usepackage{tcolorbox}

\usepackage[numbers]{natbib}  
\usepackage{multirow}
\usepackage{array}

\usepackage{amssymb}
\usepackage{xcolor}

\usepackage{pifont}

\usepackage{hyperref}   

\usepackage{hyperref}
\hypersetup{
    colorlinks=true,
    linkcolor=blue,
    filecolor=blue,
    urlcolor=blue
}

\begin{document}

\title{SocratiQ: A Generative AI-Powered Learning Companion\\for Personalized Education and Broader Accessibility\\[1em]}

\author{
    Jason Jabbour$^{\dagger}$ \hspace{.25em} 
    Kai Kleinbard$^{\dagger}$ \hspace{.25em} 
    Olivia Miller \hspace{.25em} 
    Robert Haussman \hspace{.25em} 
    Vijay Janapa Reddi \\[1em]
}
\thanks{$^{\dagger}$Jason Jabbour and Kai Kleinbard contributed equally to this research.}
\affiliation{%
  \institution{Harvard University}
  \city{Boston}
  \state{Massachusetts}
  \country{United States}
}

\email{{jasonjabbour, kak2594, omiller01, rhaussman, vjreddi}@g.harvard.edu}

\begin{abstract}
Traditional educational approaches often struggle to provide personalized and interactive learning experiences on a scale. In this paper, we present SocratiQ, an AI-powered educational assistant that addresses this challenge by implementing the Socratic method through adaptive learning technologies. The system employs a novel Generative AI-based learning framework that dynamically creates personalized learning pathways based on student responses and comprehension patterns. We provide an account of our integration methodology, system architecture, and evaluation framework, along with the technical and pedagogical challenges encountered during implementation and our solutions. Although our implementation focuses on machine learning systems education, the integration approaches we present can inform similar efforts across STEM fields. Through this work, our goal is to advance the understanding of how generative AI technologies can be designed and systematically incorporated into educational resources.
\end{abstract}

\begin{teaserfigure}
    \centering
    \includegraphics[width=\textwidth]{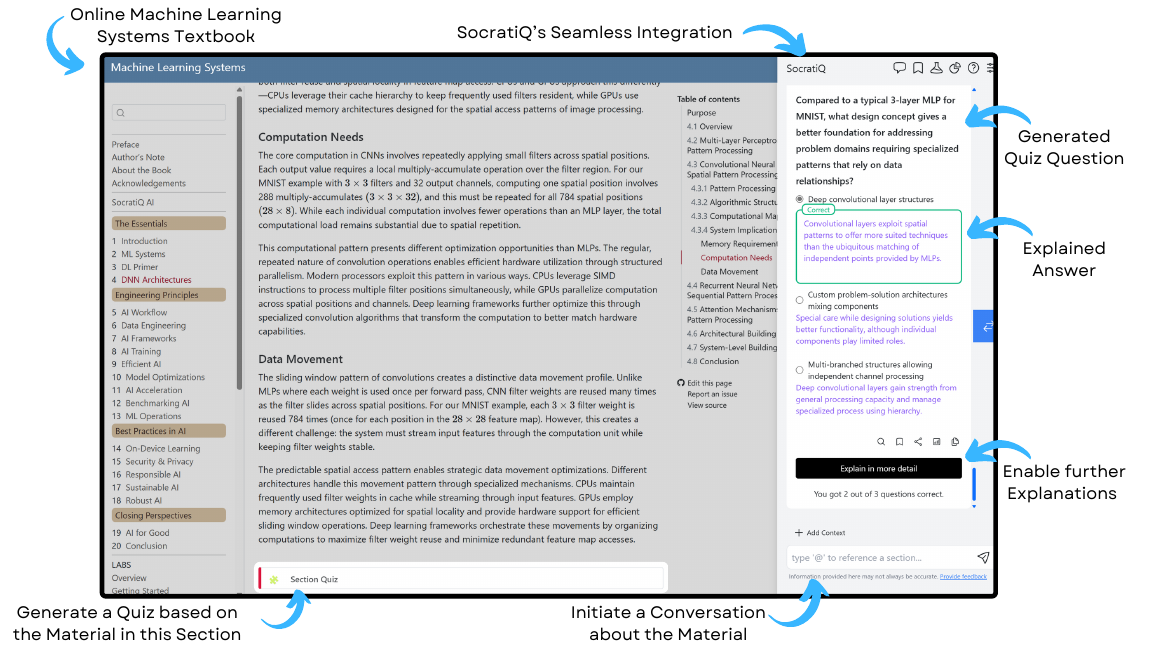}
\caption{Overview of SocratiQ's integration into the online machine learning textbook, showcasing how students can generate quiz questions and engage in natural language conversations for further explanations. You can try it at \href{https://mlsysbook.ai/}{\textcolor{blue} {https://mlsysbook.ai}}.}
    \label{fig:overview}
    \vspace{2em}
\end{teaserfigure}

\maketitle

\section{Introduction}






Education has long grappled with the challenge of meeting the unique needs of individual learners. Traditional methods often rely on standardized approaches that struggle to accommodate differences in prior knowledge, learning pace, and cognitive strengths, leaving many students without the personalized support they need to succeed. Evidence suggests that teaching styles significantly influence both student achievement \cite{aitkin1994multilevel, ebmeier1979effects} and attitudes toward subjects \cite{ebenezer1993grade}, with teaching behavior and instructional strategies playing a critical role in shaping learning outcomes \cite{opdenakker2006teacher}. Although educators may aspire to tailor instruction to every student, the scale of such efforts makes it impractical in conventional classroom settings. 

Recent advances in artificial intelligence (AI) have introduced groundbreaking possibilities in education. By leveraging the increasing sophistication of AI-driven tools, we can reimagine teaching and learning to deliver personalized and interactive educational experiences across a range of fields. Large language models (LLMs) have rapidly advanced in their ability to reason through complex problems and articulate solutions \cite{wei2022chain}. They now excel at tackling challenging subjects like mathematics, physics, chemistry, biology, and law with graduate-level rigor \cite{phan2025hle, sawada2023arb, rein2023gpqa, glazer2024frontiermath, hendrycks2021measuring}. Through fine-tuning, these models have been adapted to emulate human-like communication, becoming proficient conversationalists capable of natural dialogue and effective question answering \cite{safdari2023personality, wei2021finetuned, ziegler2019fine}. As a result, LLMs have become popular tools for learners, often serving as chatbots where students can seek clarification or explore topics.

While these tools hold significant potential to enhance education, the challenge lies in moving beyond their current ad hoc use and integrating them meaningfully into structured learning environments. Achieving this integration requires addressing gaps in delivering personalized, interactive, and course-specific educational experiences. To address this challenge, this paper introduces \textit{SocratiQ, an AI-powered learning companion} that evolves with the learner's progress. At its core is the principle of ``Generative Learning", a cognitive science framework where learners actively connect new information to prior knowledge by integrating new experiences with existing knowledge structures \cite{fiorella2016eight}. This framework emphasizes engaging cognitively through modes such as interactive dialogue (e.g., asking and answering questions). This approach contrasts with passive learning, which treats students as mere recipients of information. SocratiQ operationalizes Generative Learning through personalized explanations, adaptive assessments, and thought-provoking interactive conversations, fostering deeper comprehension, application of concepts, and higher-order thinking. 

We have integrated SocratiQ into our online \textit{Machine Learning Systems} textbook, developed at Harvard University for the CS249r course and available at \href{https://mlsysbook.ai/}{\textcolor{blue}{MLSysBook.ai}}. As a course focused on machine learning systems, this textbook is an ideal testbed for SocratiQ because it requires students to understand how different technical domains interact in ML applications, ranging from the fundamentals of neural networks to understanding their design and implementation in computer system design. Students engaging with this college-level textbook must draw on these diverse areas of prior knowledge to understand how ML systems work in practice, making standardization less effective for meaningful learning.

SocratiQ allows students to engage with interactive quizzes, receive real-time feedback, request in-depth explanations of advanced concepts, and tailor the conversation to their own level of understanding. Fully integrated into the reading experience of \href{https://mlsysbook.ai/}{\textcolor{blue}{MLSysBook.ai}} as shown in Figure \ref{fig:overview}, it creates an engaging learning environment for a textbook. The primary objectives of this integration are threefold: first, to enhance student engagement by transforming passive reading into an active, participatory experience; second, to provide personalized learning pathways that adapt to individual student needs and learning styles; and third, to provide broader accessibility to high-quality machine learning education.






We anticipate that the integration of SocratiQ into \href{https://mlsysbook.ai/}{\textcolor{blue}{MLSysBook.ai}} has the potential to transform how students interact with and learn from textbooks, particularly in complex, multidisciplinary fields. Having implemented this prototype in our CS249r course, we share our experiences and insights with the broader community. We address several key questions that are likely to interest educators, researchers, and policymakers working in AI-enhanced education.

\begin{itemize}
\item \textbf{For educators:}
\textit{What are the practical steps to implement an AI learning assistant in an online textbook or course resource?}
We share our hands-on experience integrating SocratiQ into the CS249r course textbook, focusing on the practical challenges encountered and the solutions developed. 


\item \textbf{For researchers:}
\textit{How does the integration of AI in the classroom affect learning outcomes and what insights can be gained to improve student assessment systems?}
We present a set of characteristics that focus on improving comprehension, retention, and critical thinking skills. We explore strategies for adaptive assessments to better guide students in identifying and addressing gaps in their knowledge.
\item \textbf{For policymakers:}
\textit{How can AI be integrated into classroom settings while ensuring that students remain active participants in the learning process?}
We highlight that AI tools, such as SocratiQ, complement rather than replace critical learning activities. We demonstrate how AI can guide students to actively engage with the material, ensuring mastery of concepts rather than passive reliance on AI-generated answers.

\end{itemize}

The remainder of this paper is organized as follows. We first review related work on AI-enabled educational tools, examining both their capabilities and limitations. We then compare traditional and AI-enhanced learning approaches, analyzing their respective strengths and areas for improvement. Next, we present SocratiQ's key features and system architecture, followed by detailed operational strategies for implementing such a system in practice. We conclude with an evaluation of SocratiQ's effectiveness in generating educational content and fostering student engagement, providing insights for future development of AI learning companions.


\newcommand{\neoncheck}{\textcolor[HTML]{00FF00}{\textbf{\Large\ding{51}}}} 

\newcommand{\redx}{\textcolor[HTML]{FF0000}{\textbf{\Large\ding{55}}}} 

\section{The Evolving Educational Landscape}
\label{sec:evolve}

Artificial intelligence has transformed education by reinventing traditional teaching methods. AI-driven tools improve personalization, streamline administrative tasks, and promote engagement, enriching the educational experience. However, they also pose unique challenges, such as overreliance. In this section, we trace the evolution of AI tools from early adaptive systems to modern language models, examining their strengths, limitations, and transformative impact on education, which sets the stage for SocratiQ.

\subsection{The Rise of AI-Enabled Educational Tools}

\begin{table*}[t]
\caption{Evolution of AI in education, from adaptive learning systems to large language models.}
\label{tab:ai_evolution}
\centering
\small
\begin{tabularx}{\textwidth}{>{\raggedright\arraybackslash}l|l|X|X|X}
\toprule
\textbf{Wave} & \textbf{Period} & \textbf{Representative Technologies} & \textbf{Key Capabilities} & \textbf{Impact on Education} \\
\midrule
First Wave~(Adaptive Learning) & 1990s-2010s & 
ITS (MATHia, ALEKS), Early adaptive platforms (DreamBox) & 
Rule-based personalization, Performance tracking, Automated assessments & 
Pioneered adaptive learning, Demonstrated value of personalization, Limited to structured domains. \\
\midrule
Second Wave\newline (Specialized AI) & 2010s-2020 & 
Language platforms (Duolingo, Rosetta Stone), STEM tools (Wolfram Alpha, GeoGebra), Writing assistants (Grammarly, Turnitin) & 
Task-specific AI, Interactive features, Automated feedback & 
Enhanced domain expertise, Improved engagement, Created specialized support systems. \\
\midrule
Third Wave\newline (Generative AI) & 2020-Present & 
LLMs (ChatGPT, Gemini), AI tutors (Khanmigo), Enhanced platforms (Duolingo Max) & 
Natural conversation, Context understanding, Dynamic adaptation & 
Enabled open-ended learning, Broadened accessibility, Changed student-AI interaction. \\
\bottomrule
\end{tabularx}
\end{table*}

The evolution of AI in education has progressed through several distinct phases, as summarized in Table \ref{tab:ai_evolution}. Early advances in AI laid the foundation for personalized and adaptive learning tools. Intelligent Tutoring Systems (ITS), such as Carnegie Learning's MATHia \cite{mathia}, ALEKS \cite{aleks}, and DreamBox Learning \cite{dreambox}, provided tailored instruction using techniques such as rule-based reasoning, data mining, and Bayesian networks. They offered personalized pathways for students by adapting lessons to individual progress and needs, particularly in mathematics and science education \cite{ma2014intelligent, mousavinasab2021intelligent}. Language platforms powered by AI such as Duolingo \cite{duolingo}, Rosetta Stone \cite{rosettastone}, and Babbel \cite{babbel} used Automatic Speech Recognition (ASR) to improve pronunciation and give feedback~\cite{golonka2014technologies}. 

The second wave brought specialized AI tools that targeted specific educational needs. AI-powered platforms such as Wolfram Alpha \cite{wolframalpha}, GeoGebra \cite{geogebra}, and Labster \cite{labster} supported STEM education with interactive features, self-directed learning, and collaborative opportunities \cite{huffaker2003new}. Furthermore, AI writing assistants such as Grammarly \cite{grammarly}, Turnitin \cite{turnitin} and RefME \cite{citethisforme} improved academic writing by providing grammar correction, plagiarism detection, and citation assistance~\cite{dong2021using, oneill2019stop}. However, these tools were not conversationalists; they were designed to perform specific, predefined tasks using models trained for their respective functionalities. 

The third wave, driven by advances in natural language processing (NLP), has led to the development of more versatile LLMs such as OpenAI's ChatGPT \cite{achiam2023gpt} and Google's Gemini \cite{team2023gemini}, which show strong capabilities in question answering and maintaining conversational interactions. These generative AI platforms are transforming education by offering instant answers, personalized explanations, and interactive dialogues that enhance learning and student engagement \cite{chan2023students, correia2024beyond}. The integration of generative AI technologies such as GPT-4 has significantly expanded the capabilities of educational tools. For example, Khan Academy's Khanmigo \cite{khanmigo} evolved from a platform for instructional videos and exercises into a virtual tutor powered by generative AI, offering contextual guidance and adaptive support to enhance student understanding and autonomy \cite{kshetri2023economics}. Similarly, Duolingo Max \cite{duolingomax}, leveraging GPT-4, now provides interactive role-play scenarios and contextual practice, creating realistic and immersive language learning experiences \cite{marr2023amazing}. 

Looking at this progression (Table \ref{tab:ai_evolution}), we can see how educational AI has evolved from narrow domain-specific applications to more flexible, interactive learning companions. Although early systems excelled at structured tasks like problem solving in mathematics, and specialized tools brought focused expertise to specific subjects, modern AI systems are beginning to bridge multiple domains and adapt to diverse learning contexts. This evolution reflects broader changes in educational needs, where the ability to integrate knowledge across disciplines and participate in open-ended learning has become increasingly important. 

However, while these AI systems offer powerful capabilities, their effective integration into existing educational frameworks remains a significant challenge, requiring careful consideration of how to preserve and enhance traditional teaching approaches.

\begin{table*}[ht]
\centering
\caption{A framework for understanding the AI roles that are possible in a classroom setting, along with the various design considerations, ranging from individual learning to institutional implementation, that need consideration or are required. Green checkmarks indicate features that are incorporated in the current version of SocratiQ.}
\label{tab:ai_roles_framework}
\resizebox{\textwidth}{!}{%
\begin{tabular}{llll}
\toprule
& \multicolumn{3}{c}{\textbf{AI Roles →}} \\ 
\cmidrule(lr){2-4}
\multicolumn{1}{c}{\textbf{Dimensions ↓}} &
  \textbf{Learning Companion} &
  \textbf{Teaching Assistant (AI-TA)} &
  \textbf{Assessment Coordinator} \\ 
\midrule
\multicolumn{4}{l}{\textit{Foundation \& Purpose}} \\
\textbf{Target Users} & 
  \neoncheck~Individual learners & 
  \neoncheck~ Students and instructors & 
  \neoncheck~ Administrators and faculty \\
\textbf{Primary Objective} & 
  \neoncheck~ Individual mastery of concepts & 
  \redx~Facilitate group learning & 
  \neoncheck~ Course-wide learning assessment \\
\textbf{Key Use Cases} & 
  \neoncheck~Self-paced learning, concept review   & 
  \redx~Office hours, group discussions & 
  \neoncheck~Curriculum planning, performance tracking  \\
\midrule
\multicolumn{4}{l}{\textit{Capabilities \& Features}} \\
\textbf{Generative AI Features} & 
  \neoncheck~Personalized explanations, adaptive hints & 
  \redx~Answer synthesis, discussion prompts& 
  \neoncheck~Assessment generation, feedback analysis \\
\textbf{Interactive Capabilities} & 
  \neoncheck~Real-time dialogue, concept mapping  & 
  \redx~Group facilitation, query handling  & 
  \neoncheck~Performance visualization, reporting  \\
\textbf{Personalization Level} & 
  \neoncheck~High (individual learning paths)  & 
  \redx~Medium (group-aware responses)  & 
  \redx~Low (standardized assessment)  \\
\midrule
\multicolumn{4}{l}{\textit{Technical Architecture}} \\
\textbf{Infrastructure Needs} & 
  \neoncheck~Edge computing, local processing    & 
  \redx~On-premises database  & 
  \redx~Data warehouses, analytics engines  \\
\textbf{Data Architecture} & 
   \neoncheck~Local-first, encrypted storage & 
  \redx~Distributed, session-based    & 
   \redx~Centralized, hierarchical   \\
\textbf{Integration Points} & 
  \neoncheck~E-readers  & 
  \redx~LMS plugins  & 
  \redx~Grade books, admin dashboards  \\
\midrule
\multicolumn{4}{l}{\textit{Implementation \& Deployment}} \\
\textbf{Privacy Requirements} & 
 \neoncheck~ Individual data protection  & 
   \redx~Group interaction privacy  & 
   \redx~Institutional data governance  \\
\textbf{Deployment Model} & 
  \neoncheck~Client-side focused  & 
   \neoncheck~Hybrid client-server & 
  \redx~Server-side focused   \\
\textbf{Key Challenges} & 
   \neoncheck~Context retention, personalization & 
   \redx ~Group dynamics, response timing& 
   \neoncheck~Assessment fairness, data quality \\
\midrule
\multicolumn{4}{l}{\textit{Evaluation Framework}} \\
\textbf{Learning Metrics} & 
   \neoncheck~Knowledge retention, engagement & 
   \neoncheck~Participation, query resolution & 
  \neoncheck~ Course completion, outcomes \\
\textbf{System Metrics} & 
   \neoncheck~ Response time, adaptation accuracy & 
   \neoncheck~Concurrent users, availability & 
   \neoncheck~Processing speed, data accuracy \\
\textbf{Success Indicators} & 
   \neoncheck~Student progress, satisfaction  & 
   \neoncheck~Reduced instructor load, accessibility  & 
   \neoncheck~Curriculum effectiveness, ROI \\
\bottomrule
\end{tabular}%
}
\end{table*}

\subsection{Balancing AI \& Human-Centered Education}

The evolution of education has seen significant advances in both traditional and AI-assisted learning models, each contributing distinct benefits to the development of well-rounded students. Traditional educational models offer irreplaceable advantages in the cultivation of interpersonal relationships and a vibrant sense of community among students. Collaborative activities, such as group projects and discussions, facilitate learning from diverse perspectives, thus enhancing critical thinking and problem solving skills.

In addition, experiential learning opportunities, including research projects and laboratory activities, establish crucial connections between classroom instruction and real-world applications, thus improving knowledge retention and comprehension. In particular, the presence of human instructors provides essential mentorship, individualized guidance, and support for the development of vital soft skills such as communication and presentation abilities - skills that remain critical for professional development.

Concurrently, AI's capabilities have revamped educational possibilities through unprecedented personalization and scalability. The capacity of AI to deliver customized learning experiences represents one of its most significant contributions to education: generating customized materials, creating adaptive quizzes, and providing writing assistance that improves student engagement and study time \cite{pesovski2024generative, chen2020artificial}. These tools have demonstrated particular efficacy in STEM fields, where computer-assisted instruction significantly improves performance compared to traditional methods \cite{morgil2005traditional}. 

The ability of AI to provide immediate feedback enables students to identify strengths and areas for improvement in real time, while its automation of routine tasks such as grade allows instructors to focus on meaningful student interactions \cite{chen2020artificial}. Importantly, AI contributes to the democratization of education by offering high-quality learning resources to students who might otherwise lack access to traditional educational opportunities \cite{baum2019human}.

Despite the complementary potential of traditional and AI enhanced approaches, their integration presents challenges that require careful consideration. Traditional learning models are constrained by temporal, spatial, and resource limitations, which impediments the scalability of personalized attention, particularly in large educational settings. In contrast, AI-assisted models introduce their own complexities, ranging from privacy and security concerns arising from extensive data collection to implementation barriers such as limited understanding of data-driven systems and questions of data sovereignty \cite{renz2020prerequisites}. A key challenge lies in mitigating the potential erosion of human connections; while AI excels in content delivery, it cannot replicate the empathetic and nuanced support that human educators provide for students' emotional and psychological needs \cite{dhanapal2024impact}. Furthermore, there exists a risk of overreliance on AI tools, which could potentially impede the development of critical thinking skills if students default to AI-generated answers rather than engaging deeply with the material \cite{al2024exploring}. Consequently, we must navigate the delicate balance between leveraging AI's capabilities and preserving the essential human elements of learning.

The creation of an effective educational environment requires the thoughtful integration of traditional and AI-assisted approaches while preserving their respective strengths. In traditional education, various roles collectively improve the learning experience, from peer-to-peer support among classmates to teaching assistants that bridge gaps between students and instructors, to course coordinators who supervise curriculum quality (Table \ref{tab:ai_roles_framework}). AI systems should be designed to complement rather than replace these roles, addressing specific gaps such as providing personalized support in large courses or augmenting peer learning where additional assistance is required. The sollution lies in leveraging AI's capabilities for personalization and efficiency while maintaining the human connections and community support essential for meaningful learning experiences. This balanced approach can foster both immediate engagement and long-term development: while AI provides immediate feedback and adaptive pathways tailored to student skill levels, traditional methods build self-efficacy through collaborative projects and substantive human interactions.

The rapid evolution of AI technology emphasizes the need for systematic frameworks to guide this integration. Such frameworks must carefully balance AI's capabilities with traditional pedagogical strengths while addressing the challenges of implementation. This objective motivated the development of our SocratiQ AI learning companion system, which demonstrates how AI can enhance rather than replace the human elements central to effective education.

\section{SocratiQ System Design}
\label{sec:design}

In this section, we introduce SocratiQ as a learning companion designed for an online textbook on machine learning systems, focusing on how it addresses the specific challenges of teaching such a complex subject. SocratiQ addresses the unique challenges inherent in teaching complex and multidisciplinary subjects by integrating advanced AI capabilities with traditional pedagogical approaches.

The design of SocratiQ is based on the framework presented in Table~\ref{tab:ai_roles_framework}, with a primary focus on fulfilling the role of the learning companion. This approach allows us to leverage AI's strengths in personalization and scalability while preserving the irreplaceable aspects of human-centered education. We identified four key features critical to an effective AI learning companion: personalized explanations, adaptive assessments, bounded learning, and gamification. These features are carefully designed to address the specific challenges of teaching machine learning systems, a field that requires students to synthesize knowledge from diverse areas such as algorithms, linear algebra, and computer architecture. 

In the following subsections, we explore each of these features in depth. We discuss the rationale behind their inclusion (based on our experience in teaching CS249r) and detail their structure and implementation. This  overview hopes to provide educators with a practical framework for incorporating similar AI-enhanced elements into their own courses, regardless of the subject matter.

\begin{figure}[t!]
    \centering
    \includegraphics[width=.85\linewidth]{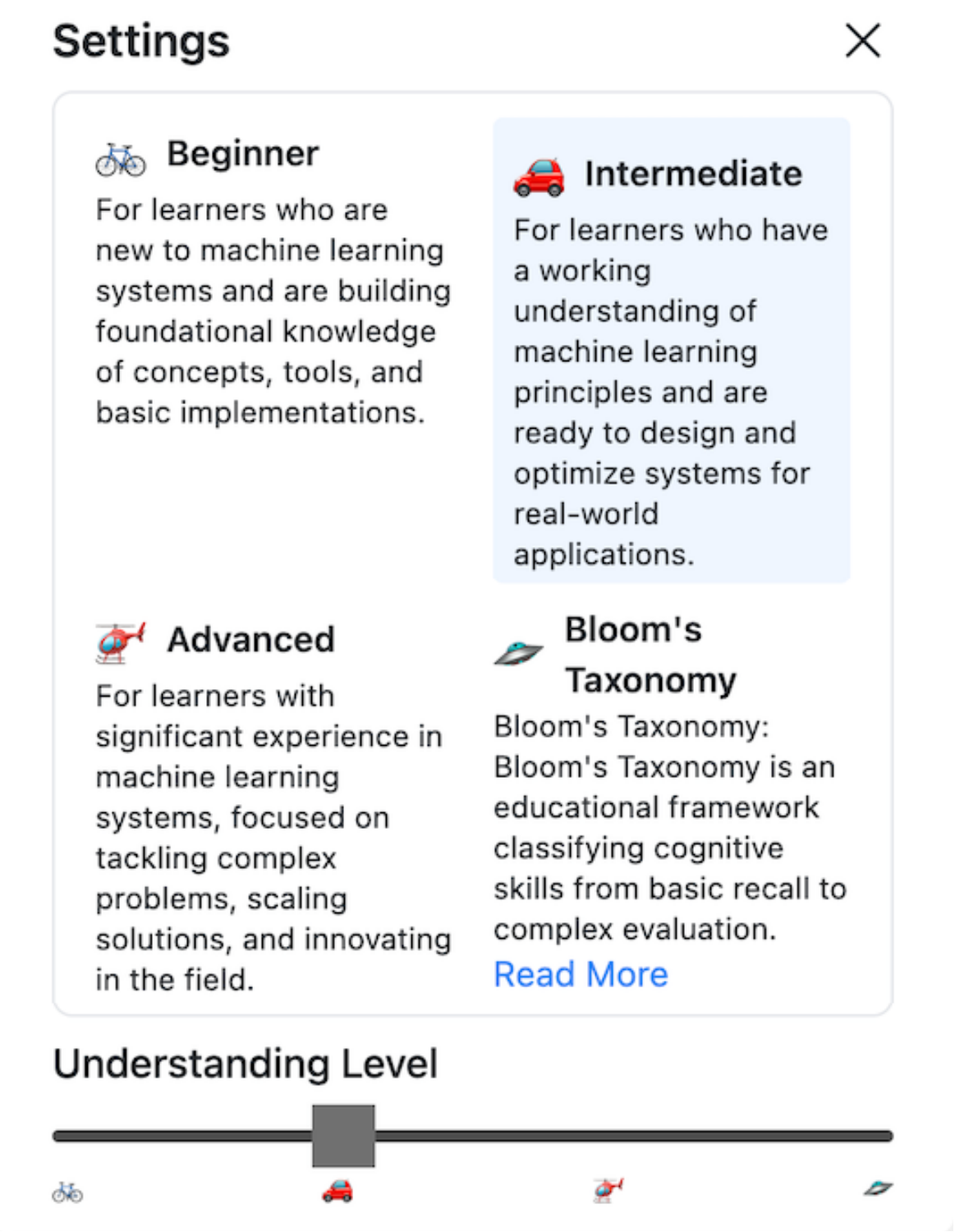}
    \caption{Students can dynamically adjust the academic level to match their learning preferences.}
    \label{fig:difficulty-level-slider}
    \vspace{-2em}
\end{figure}

\subsection{Personalized Explanations}
\label{personalized_expl}

Students enter courses with varying levels of prior knowledge, a challenge that is particularly evident in advanced college-level curricula that integrate concepts from multiple disciplines. This variation is especially pronounced in certain types of classes, which require thorough understanding across several foundational areas. 

The CS249r Machine Learning Systems class at Harvard University exemplifies this challenge, illustrating the complex interplay of knowledge required in this field. The course demands not only theoretical knowledge, but also practical application skills. Students are required to design and implement efficient and scalable machine learning systems that can handle large datasets and complex models. This requires a deep understanding of the principles of system design and the ability to optimize algorithms for various hardware configurations. For instance, they must grasp the underlying algorithms of machine learning, such as backpropagation and stochastic gradient descent, while simultaneously comprehending linear algebra principles, including matrix multiplication used in neural networks. Furthermore, they need to possess familiarity with computer architecture, understanding how numerical representations are handled at a low level and how different hardware architectures impact machine learning model quality and performance.

The breadth of these prerequisites results in a student body with widely varying degrees of expertise across these areas. Some students may excel in theoretical machine learning concepts but struggle with systems-level implementation. Others might have a strong background in computer architecture but find the intricacies of advanced machine learning algorithms challenging. This disparity in background knowledge often leads to significant gaps that can impede a student's ability to fully engage with the material.

The challenge in CS249r lies not just in teaching individual concepts, but in helping students bridge these diverse areas of knowledge, creating a holistic understanding of machine learning systems that spans from theoretical foundations to practical, efficient implementations. This multifaceted nature of the course highlights the complexity of teaching advanced, interdisciplinary subjects and the need for adaptive, comprehensive educational approaches.

In such scenarios, a one-size-fits-all approach to explanation— delivering content in a single, standardized manner—is often ineffective. Effective comprehension requires adapting explanations to align with a student’s background knowledge and addressing specific areas of difficulty. Adjusting the complexity and depth of explanations enables learners to integrate new information with their existing understanding, improving retention and engagement.

To address this, SocratiQ provides four difficulty levels that allow the AI learning companion to tailor its explanations to the student’s current level of understanding. The difficulty range spans from Easy to Expert, aligning content with learners' cognitive abilities. At the Expert level, SocratiQ employs Bloom’s Taxonomy to pose higher-order thinking style questions. This personalization is achieved through a user interface that enables learners to adjust their knowledge level using a slider menu as displayed in Figure \ref{fig:difficulty-level-slider}. The difficulty levels are defined as follows:

\begin{enumerate}
    \item \textbf{Beginner}: Focus on foundational concepts, definitions, and straightforward applications in machine learning systems, suitable for learners with little to no prior knowledge.
    \item \textbf{Intermediate}: Emphasize problem-solving, system design, and practical implementations, targeting learners with a basic understanding of machine learning principles.
    \item \textbf{Advanced}: Challenge learners to analyze, innovate, and optimize complex machine learning systems, requiring deep expertise and a holistic grasp of advanced techniques.
    \item \textbf{Expert (Bloom's Taxonomy)}: Create responses that progress through Bloom's levels: remember, understand, apply, analyze, evaluate, and create. Guide my learning.
\end{enumerate}

These difficulty levels are implemented as system prompts provided to the language model, enabling users to adjust the platform to align with their learning needs. The corresponding prompts for each difficulty level, simplified here for brevity, are as follows:

\vspace{1em}

\definecolor{lightlavender}{RGB}{175,175,250} 
\definecolor{mintgreen}{RGB}{152,251,152} 
\definecolor{palepeach}{RGB}{255,239,150} 
\definecolor{powderblue}{RGB}{120,224,230} 

\textbf{Beginner Prompt:}
\vspace{-.7em}
\begin{tcolorbox}[colback=powderblue!20,colframe=gray!50,width=\linewidth]
 ``You are conversing with a Beginner learner: Focus on foundational concepts, definitions, and straightforward applications in machine learning systems, suitable for learners with little to no prior knowledge.''
\end{tcolorbox}

\textbf{Intermediate Prompt:}
\vspace{-.7em}
\begin{tcolorbox}[colback=powderblue!20,colframe=gray!50,width=\linewidth]
 ``You are conversing with an Intermediate learner: Emphasize problem-solving, system design, and practical implementations, targeting learners with a basic understanding of machine learning principles.''
\end{tcolorbox}

\textbf{Advanced Prompt:}
\vspace{-.7em}
\begin{tcolorbox}[colback=powderblue!20,colframe=gray!50,width=\linewidth]
 ``You are conversing with an Advanced learner: Challenge learners to analyze, innovate, and optimize complex machine learning systems, requiring deep expertise and a holistic grasp of advanced techniques.''
\end{tcolorbox}

\textbf{Expert Prompt:}
\vspace{-.7em}
\begin{tcolorbox}[colback=powderblue!20,colframe=gray!50,width=\linewidth]
 ``You are an expert ML teacher using Bloom's Taxonomy: Create responses that progress through Bloom's levels: remember, understand, apply, analyze, evaluate, and create. Guide my learning.''
\end{tcolorbox}

While these prompts are tailored to the specific challenges of teaching machine learning systems, they provide us with a model that can be easily customized for other domains outside of this subject. The core principle of identifying and addressing the diverse knowledge backgrounds of students, bridging interdisciplinary gaps, and creating adaptive learning experiences is universally applicable. Therefore, by adapting this framework, educators in various disciplines can create more effective learning environments that cater to the diverse needs of their students.

\subsection{Adaptive Assessments}
\label{adaptive_assessments}

Traditional static assessments often struggle to capture whether a student has truly understood the material or provide opportunities to address gaps in their knowledge. Although giving a student one chance to answer a question provides some indication of their comprehension, it does little to help them identify and address areas of misunderstanding. In addition, instructors are typically limited in their ability to create a wide variety of assessments, making it challenging to offer the repeated and varied testing opportunities necessary for comprehensive learning.

In CS249r, for example, students tackle complex tasks such as optimizing neural networks for GPU execution. A traditional assessment might ask all students to implement this optimization, regardless of their prior knowledge. However, this approach does not address the diverse starting points of students. For instance, a student with strong machine learning theory but limited GPU knowledge would struggle differently than one with extensive systems experience but less algorithmic understanding. An adaptive assessment could initially gauge each student's strengths, then provide tailored subtasks and resources, ensuring a more effective and personalized learning experience for all.

An effective alternative is an adaptive assessment approach that provides students with multiple opportunities to test their knowledge, focus on areas of weakness, and revisit the material in a meaningful way \cite{hattie2007, shute2008}. This approach is particularly valuable in courses like CS249r, where students' diverse backgrounds requires personalized learning paths. This process should not rely on repeated exposure to the same questions, as that would merely test rote memorization or random chance rather than true understanding. Instead, adaptive assessments should dynamically adjust to focus on the concepts the student has yet to master, ensuring a more targeted and effective learning experience.

To realize such an adaptive assessment approach, we leverage an LLM to dynamically generate assessments, providing learners with frequent opportunities to check their understanding and track their progress. This feature is integrated into our online machine learning systems textbook, which includes chapters on a variety of topics such as AI Acceleration, Model Optimizations, and Machine Learning Operations. Each chapter contains extensive content with which students are expected to engage and the information in a chapter is divided into sections like any typical textbook.

When a student visits the website, the text in the chapter is indexed into discrete sections based on HTML heading tags (H1, H2, etc.), and the corresponding text is saved in memory. A button is also added after each section. When a button is clicked, indicating that the learner would like to test their understanding of the material, the associated text is passed to the LLM with the following instructions:

\vspace{1em}
\textbf{Quiz Generation Prompt:}
\vspace{-.7em}
\begin{tcolorbox}[colback=powderblue!20,colframe=gray!50,width=\linewidth]
\footnotesize
\begin{verbatim}
Create a quiz from a CHAPTER SECTION.
The quiz should have 3 questions in JSON format:
- Q1 \& Q2: Directly related to the quote's content.
- Q3: Requires deeper understanding.
Use this JSON template, modifying it as needed:
{"questions": [
    {"question": "Q1 here?",
      "answers": [
        {"text": "A1", "correct": true/false, 
        "explanation": "explanation"},
        {"text": "A2", "correct": false, 
        "explanation": "explanation"},
        {"text": "A3", "correct": false, 
        "explanation": "explanation"},
      ]},
    {"question": "Q2 here?", "answers": [/* options */]},
    {"question": "Q3 here?", "answers": [/* options */]}
  ]}
QUOTE: ${<insert quote here>}
CHAPTER SECTION ${<insert name and section number of chapter>}$
\end{verbatim}
\normalsize 
\end{tcolorbox}

The LLM returns a JSON object containing the questions, multiple-choice options, and detailed answer explanations embedded in the response. This JSON object is rendered for the learner to interact with, providing them with immediate feedback on their quiz. 

\begin{figure}
    \centering
    \includegraphics[width=1\linewidth]{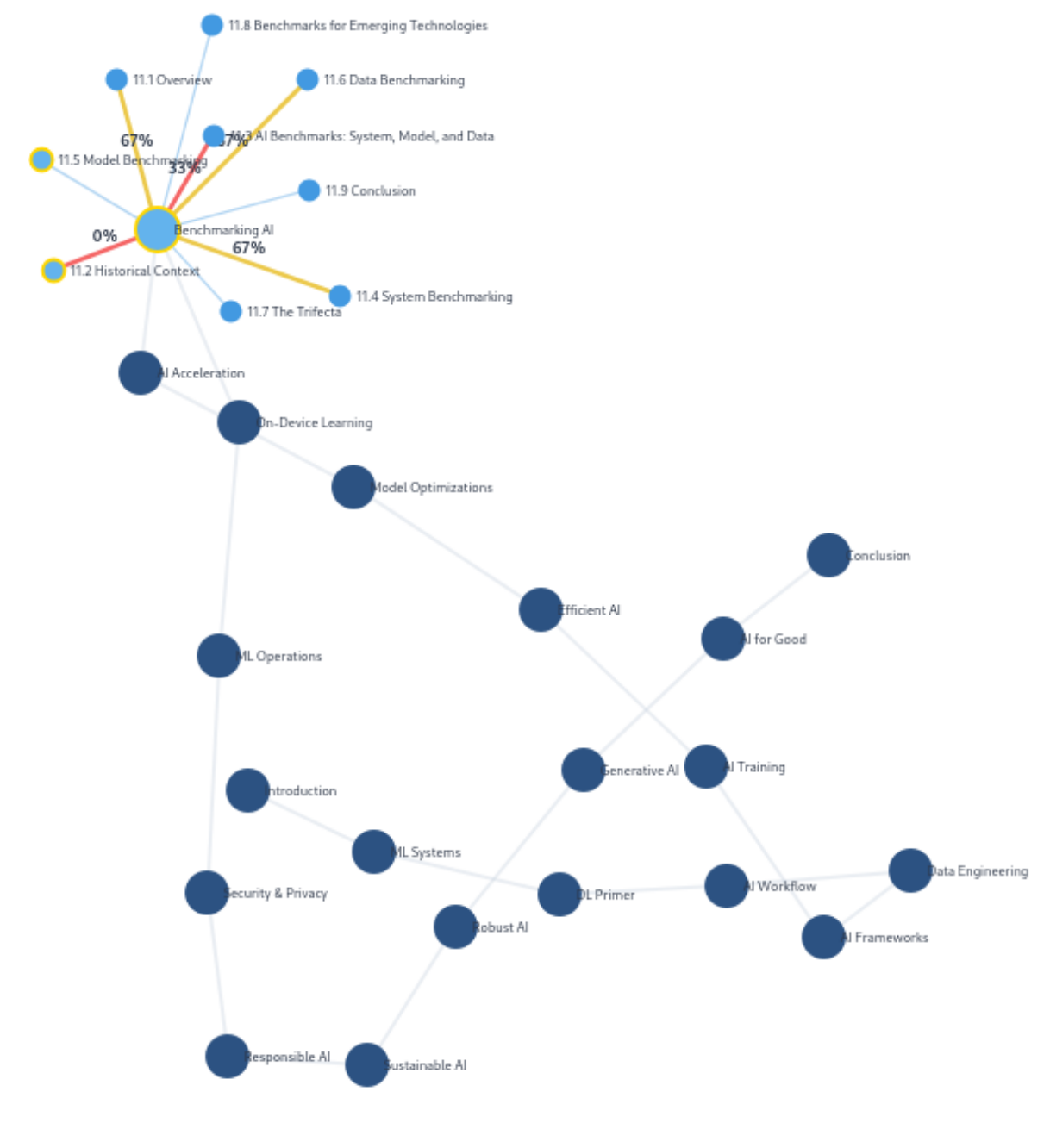}
    \caption{As students progress through the textbook, a knowledge graph is built, tracking their reading progress and quiz performance for each section.}
    \label{fig:knowledge-graph}
    \vspace{-1em}
\end{figure}

As students progress through the sections of the textbook and complete quizzes, a knowledge graph is constructed to outline various topics and subtopics relevant to their coursework, as shown in Figure \ref{fig:knowledge-graph}. This graph tracks the content a student has engaged with, their quiz proficiency, and the types of questions they have answered incorrectly. The knowledge graph provides students with an interactive tool to select sections for assessment, enabling them to focus on areas of strength or weakness. Students can also use the graph to seek AI-driven advice on their learning, identify patterns, and receive guidance on what to study next.


\subsection{Bounded Learning}

Large language models are trained on vast corpora of text that span much of the Internet, equipping them with strong reasoning and comprehension capabilities \cite{brown2020language}. However, the specific datasets used during training are often unknown to the public, and the completeness or accuracy of the information they generate cannot be guaranteed. Thus, relying solely on the pre-trained knowledge of the language model introduces the risk of generating content that is inconsistent or not related to the course content curated.

Consider a scenario in CS249r where students are learning about tensor core optimizations for GPU acceleration. A general LLM might provide broad information about GPUs and tensors, but it may not capture the specific techniques and best practices taught in the course. For instance, when asked about optimizing matrix multiplication on tensor cores, the LLM might suggest generic CUDA programming tips rather than the specific tiling and data layout strategies covered in the class lectures and textbook.

Using AI's unbounded knowledge as a tool may not fully leverage the advantages of the carefully curated content of our textbook. Although pre-trained knowledge of the language model offers valuable information, it is essential to balance this capability with a focus on curated learning resources that were specifically designed for the course. This approach ensures that learners benefit from the accuracy, depth, and coherence of the primary textbook while still having access to broader supplementary information when needed.

Our solution emphasizes textbook material as the primary focus of the language model through in-context prompts, while retaining access to its broader pre-trained knowledge. These curated prompts explicitly bind the LLM to the information in the textbook, providing a prior that biases the model to draw primarily from curated content while still allowing it to explore related concepts as needed. In the tensor core optimization example, this means the LLM would prioritize information from the course materials about specific CUDA programming patterns for tensor cores, while still being able to provide general context about GPU architectures.

To determine which information to include within the in-context prompt, we quickly extract text relevant to any user query. To achieve this efficiently, we use Algorithm~\ref{alg:fuzzy-paragraph-matching}, which identifies similar paragraphs using fingerprinting and Levenshtein distance.

\begin{algorithm}[t!]
\caption{Fuzzy Paragraph Matching Overview}
\label{alg:fuzzy-paragraph-matching}
\begin{algorithmic}[1]
\REQUIRE Query \( Q \), precomputed fingerprints \( \text{textMap} = \{(f_i, \text{id}_i, T_i)\} \)
\ENSURE Top \( k \) most similar paragraphs to \( Q \)
\STATE \textbf{Text Preprocessing:} Prepare \( T \) by removing punctuation and converting to lowercase.
\STATE \textbf{Fingerprinting:} Compute the fingerprint for \( T \) as the average ASCII value of its characters:
\[
\text{fingerprint}(T) = \frac{\sum_{c \in T} \text{ascii}(c)}{|T|}
\]
\STATE \textbf{Precomputed Fingerprint Map:} Store fingerprints \( f_i \) for paragraphs \( T_i \):
\[
\text{textMap} = \{(f_i, \text{id}_i, T_i)\}
\]
\STATE \textbf{Finding Candidate Paragraphs:} Given a query text \(Q\) with fingerprint \(f_Q\), perform a binary search on \(\text{textMap}\) to find the closest fingerprints. Select neighboring fingerprints as candidates:
\[
\text{candidates}(Q) = \text{findClosest}(f_Q, \text{textMap})
\]
\STATE \textbf{Refined Similarity (Chunked Levenshtein Distance):} For each candidate \(C\), calculate the Levenshtein distance to \(Q\), representing the minimum edits to transform one string into another. Compute similarity as:
\[
\text{similarity}(Q, C) = 1 - \frac{\text{Levenshtein}(Q, C)}{\max(|Q|, |C|)}
\]
\STATE \textbf{Ranking and Selection:} Rank candidates by similarity and return the top \( k \).
\end{algorithmic}
\end{algorithm}





















Each page that a student visits for the first time undergoes pre-computation, where every diagram (e.g., charts, figures) and paragraph (\texttt{<p>} element in HTML) is assigned a unique fingerprint attribute. This allows us to efficiently find the most relevant paragraphs for any query based on these fingerprints.

By balancing the curated textbook content with the supplementary knowledge from the language model’s training data, this approach ensures that learners receive accurate, focused, and contextually relevant information \cite{wenger1998, bransford2000}.

\begin{figure}[t!]
    \centering
    \includegraphics[width=1.0\linewidth]{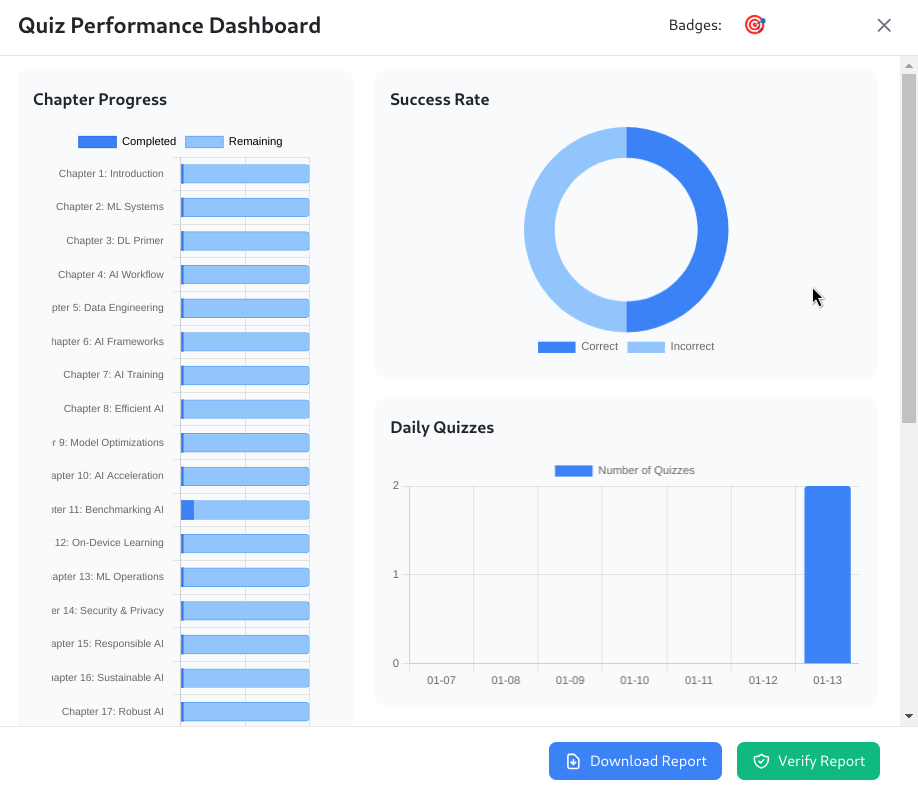}
    \caption{Dashboard showcasing gamification elements, including badges for quiz streaks and progress bars, to enhance engagement and provide learning insights.}
    \label{fig:gamification-dashboard}
\end{figure}

\begin{figure}[t]
    \centering
    \includegraphics[width=1.0\linewidth]{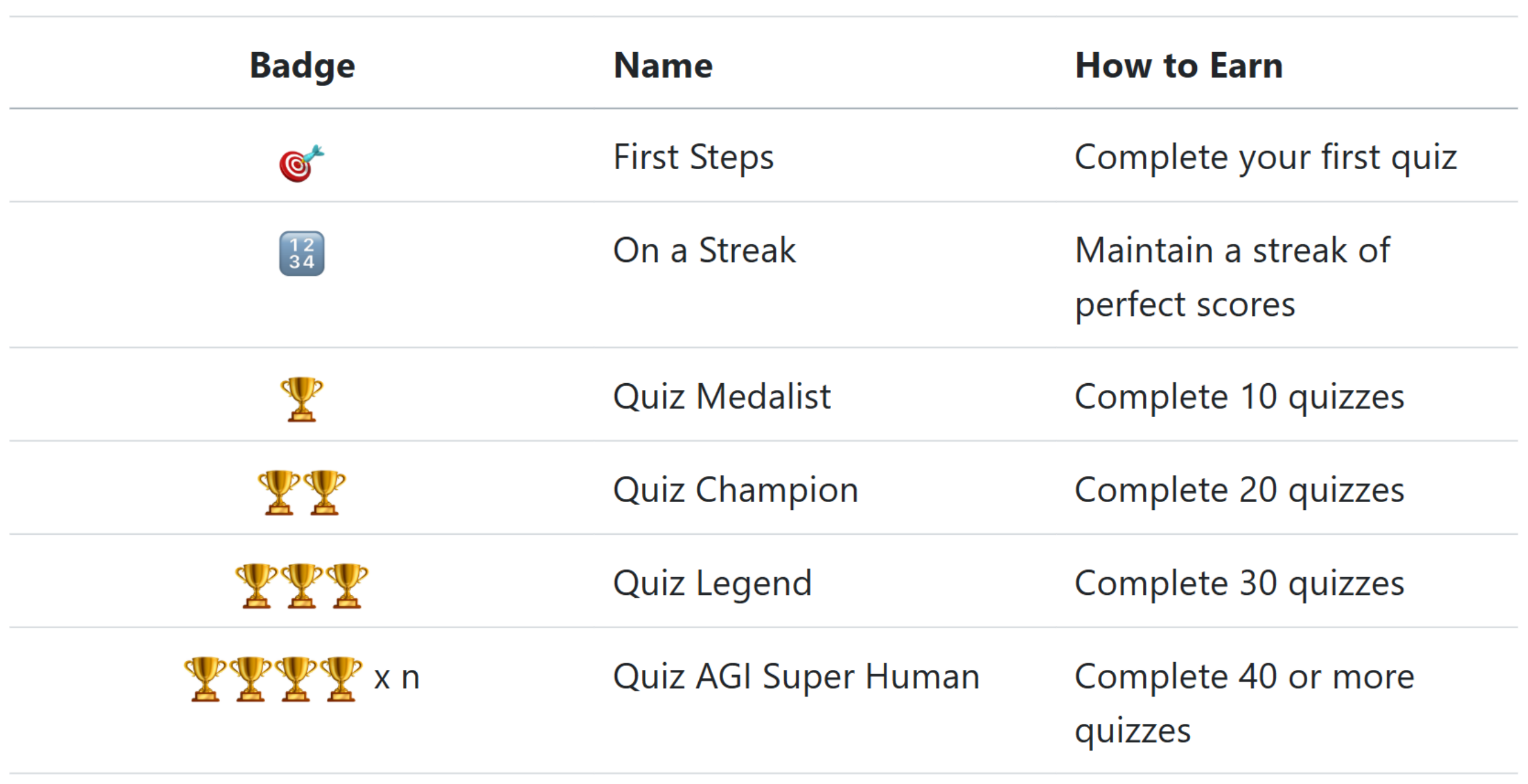}
    \caption{Overview of achievement badges students can earn through quizzes.}
    \label{fig:badges}
\end{figure}

\subsection{Gamification}

An effective AI learning companion should incorporate features that actively engage students and sustain their motivation throughout the learning process. Gamification has been shown to be an effective strategy for increasing enjoyment in educational contexts \cite{looyestyn2017does}. Research has also highlighted the mass appeal of gamification in stimulating learner motivation and promoting active engagement \cite{zainuddin2020impact}. With these elements, educators can create  engaging learning experiences that encourage both participation and persistence. 

\begin{figure}[t!]
    \centering
    \includegraphics[width=1.0\linewidth]{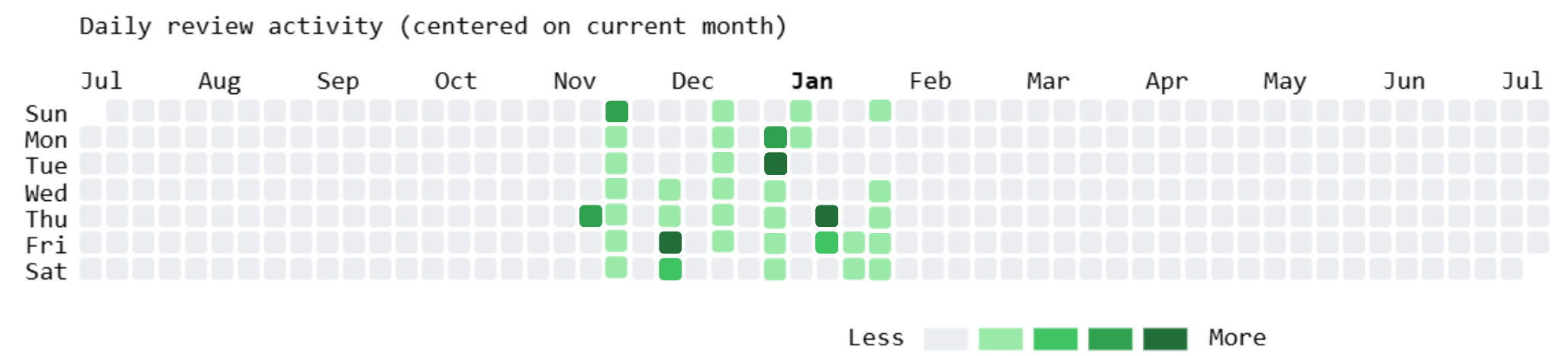}
    \caption{Socratiq displays a visualization of the frequencies of one's engagement with the app, inspired by the Github Contribution heatmap.}
    \label{fig:heatmap}
\end{figure}

SocratiQ uses gamification and progress tracking to enhance user engagement and provide opportunities for self-assessment. We achieve this by implementing five features designed to motivate learners:



\begin{enumerate}
\item \textbf{Progress}: Tracks the learner's completion of sections within a chapter as shown in Figure \ref{fig:gamification-dashboard}. The number of required sections to pass a chapter and the threshold score for a quiz are parameters set by the administrator.
\item \textbf{Streaks}: The number of consecutive days a user has taken a quiz.
\item \textbf{Passing quiz attempts}: The total number of quiz attempts a user has passed.
\item \textbf{Badges}: Users earn badges for every $c$ number of quiz attempts that pass, where $c$ is a parameter chosen by the administrator, as shown in Figure \ref{fig:badges}.
\item \textbf{Engagement Heatmap}: Visualizes user activity, with darker squares for more engagement as shown in Figure \ref{fig:heatmap}. 
\end{enumerate}
These elements provide users with a sense of accomplishment and motivation to continue using the platform.

\section{SocratiQ Implementation}

\begin{figure}[t]
    \centering
    \includegraphics[width=1\linewidth]{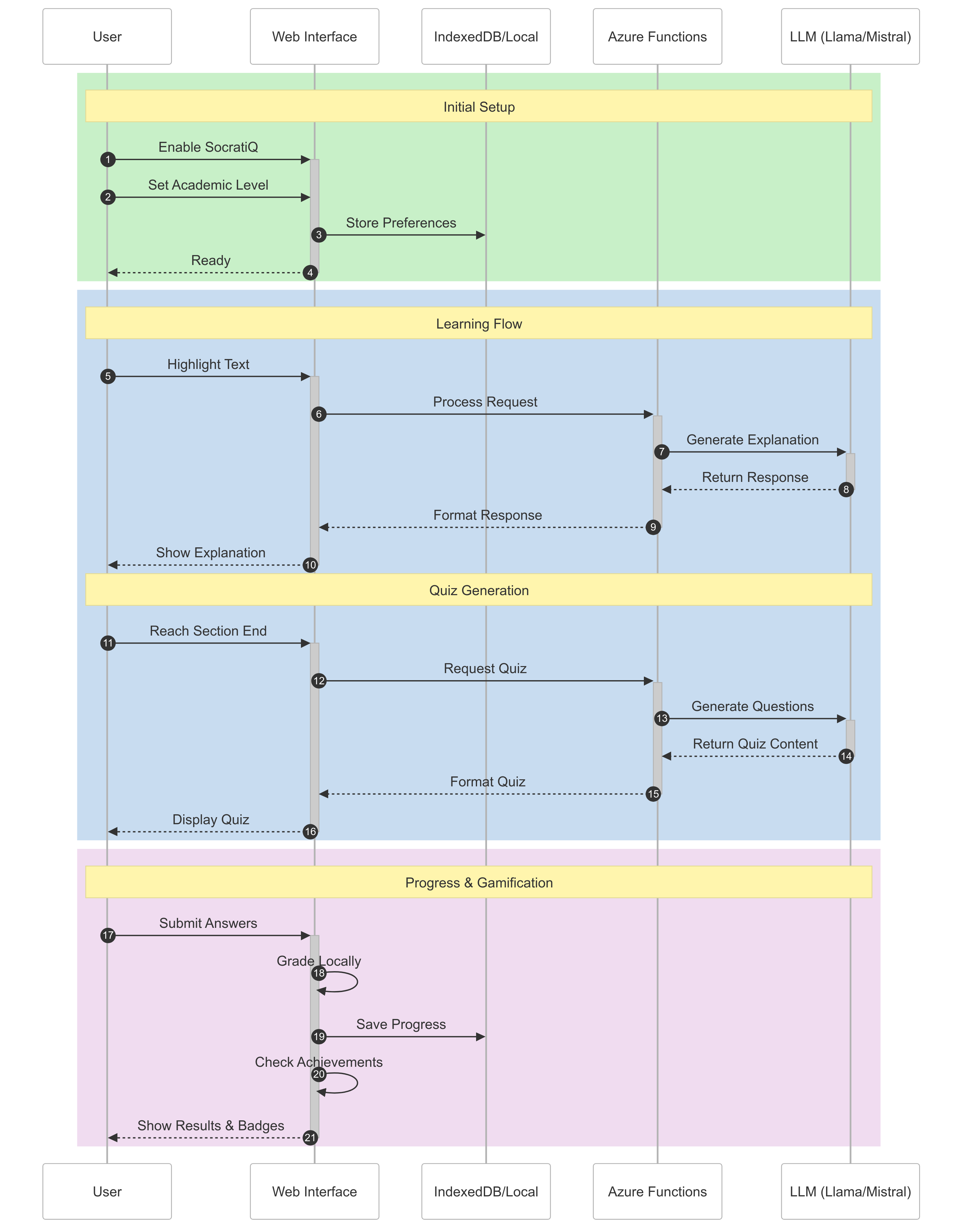}
    \caption{SocratiQ system architecture.}
    \label{fig:sys_arch}
\end{figure}

The implementation of the SocratiQ system can be organized into four main stages: Initial Setup, Learning Flow, Quiz Generation, and Progress \& Gamification as illustrated in Figure \ref{fig:sys_arch}. These stages encapsulate the key phases of user interaction and system functionality. Each stage integrates multiple components, spanning a web interface, local database, Azure, and a language model. The following sections detail the purpose and implementation of each stage, highlighting their roles within the overall architecture.

\subsection{Initial Setup}

The SocratiQ platform is implemented as a client-side application, delivered as a single JavaScript file. It is designed to integrate into any webpage, making it a versatile solution to embed AI-driven learning tools on various educational platforms. When the user enables SocratiQ, the JavaScript code injects itself into the Shadow DOM. This creates an isolated environment, ensuring that SocratiQ's styles and functionality remain independent of the host webpage. Using Shadow DOM, SocratiQ can integrate into any online environment without disrupting existing website elements or being affected by external styles or scripts. During the initialization process, the JavaScript code extracts text from the webpage, in our case, chapters from our online machine learning systems textbook. Indexes the content into discrete sections and dynamically injects quiz buttons after each section of the textbook chapter. 

Once the system is enabled, the user can personalize their academic level using a slider, as described in Section \ref{personalized_expl}. The platform stores these preferences locally in the web browser's built-in database, IndexedDB, ensuring both privacy and fast access without relying on server-side storage. Once this setup is complete, the SocratiQ learning companion is fully operational and ready to deliver a personalized and engaging learning experience.

\subsection{Learning Flow}

In our use case of an online textbook, the AI learning companion comes to life when the student simply highlights text on the page that they need clarification on. This intuitive interaction keeps the student engaged with the material, seamlessly integrating the companion into the learning process without redirecting them to another page or disrupting their focus. For instance, in CS249r, a student might highlight a passage about cache coherence protocols in multicore systems, seeking a more detailed explanation.

When a student highlights a section of text, the selected content is sent through an Azure Function. Our architecture is designed to be serverless to minimize complexity and relieve instructors of the need to manage infrastructure. Azure Functions automatically handle server provisioning, scaling, and maintenance, providing a highly scalable and efficient solution for processing requests. This design ensures that the system can handle multiple simultaneous user interactions without requiring dedicated server resources.

The Azure Function forwards the request to a language model via an API call. Using the stored preferences set during the initial setup (e.g. academic level), the language model generates an explanation tailored to the learner's needs. In our CS249r example, the model might provide a more in-depth explanation of how cache coherence protocols maintain consistent shared memory state across multiple cores, adjusting the complexity based on the student's background.

Note that these language models are hosted in the cloud and that we access them via API calls rather than hosting the models on our own servers. While this offloading reduces the complexity of maintaining high-performance infrastructure, it incurs a cost per API call, which will be discussed further in Section \ref{cost_analysis}. Once the language model returns the generated explanation, it is formatted for readability and displayed to the student.

\subsection{Quiz Generation}

At the end of a chapter section, students can select the embedded quiz button to initiate a quiz. For example, in CS249r, after studying a section on GPU architecture and CUDA programming, a student might choose to test their understanding using a generated quiz. A similar pipeline is followed as when a student highlights text: the quiz request is sent through an Azure function.

The response is then formatted and displayed to the user. A key aspect of this process is the use of the prompt structure described in Section \ref{adaptive_assessments}, which ensures that the language model generates questions relevant to the selected content, at the appropriate academic level, and in the correct format. In the CS249r course, this might result in questions about different quantization techniques (such as post-training quantization or quantization-aware training), the trade-offs between model size and accuracy in pruning, or specific strategies for structured and unstructured pruning, all tailored to the student's current level of understanding.

The amount of information sent to the language model for generating questions is constrained by token limits, and the strategies used to optimize this process are discussed in Section \ref{optimizations}. These optimizations are particularly crucial for complex topics in CS249r, where a concise yet comprehensive context is necessary to generate meaningful questions about advanced concepts such as the impact of quantization on different neural network architectures or the iterative process of pruning and fine-tuning.

\subsection{Progress \& Gamification}

Upon completing a quiz, the student's grade is saved in the browser's IndexedDB, which ensures that their performance data remain accessible without requiring server-side storage. This local storage approach also records detailed quiz statistics, current progress, and recent learning streaks, enabling comprehensive tracking of user engagement and achievement.

For CS249r students, this might include specific metrics on their understanding of model quantization techniques or pruning strategies. For instance, the system could track a student's proficiency in post-training quantization versus quantization-aware training, or their ability to analyze the trade-offs between model size reduction and accuracy preservation in various pruning scenarios.

These statistics are combined with gamification elements to provide students with a clear and motivating display of their progress. Beyond simple badges, the system incorporates challenges tailored to each student's learning trajectory. For example, a student who excels in theoretical aspects of model optimization might be challenged to apply these concepts in practical scenarios, such as optimizing a machine learning model for edge device deployment.

\section{Operationalizing SocratiQ}

This section provides insights and practical considerations for operationalizing the SocratiQ AI learning companion. We discuss various language model options, practical optimizations for handling API call limitations, cost analysis, and how the system can be integrated into the broader structure of a course.







\subsection{Language Models}

We use Groq \cite{groqcloud}, a cloud-based service, to run inference on language models for SocratiQ. It offers several key advantages: fast inference times, needed for maintaining real-time interactivity in our educational platform; elimination of the need for local deployment, avoiding the substantial costs and maintenance associated with server-grade GPUs; and scalability to handle varying loads without infrastructure management on our part. However, Groq, much like any other cloud-based service, imposes specific API call limitations, including constraints on Requests per Minute, Requests per Day, Tokens per Minute, and Tokens per Day.

A unique challenge we face is that our CS249r online textbook is open to the public as it is a website; it has been visited by over 30,000 users in the last 6 months alone. Therefore, it can potentially generate a large volume of requests from users around the world. This open access policy, aligned with our goal of democratizing education, requires a flexible and cost-effective approach to handling model inference. To address this, we use a combination of language models, including Mixstral-8x7b, Gemma 7b, and LLama 3.2, as our primary AI services. By distributing requests across these models, we optimize performance while staying within the free-tier's limits.

In the event of Groq unavailability or when demand exceeds the free tier limits, we rely on other services such as Google Gemini as a backup service to ensure uninterrupted functionality. This multi-model, multi-service strategy not only provides resilience but also allows us to leverage the strengths of different models for various educational tasks while managing costs effectively.

Note, however, that when SocratiQ is used within a controlled classroom setting, we have the option to switch to more dedicated services with higher capacity and fewer restrictions. However, for the public-facing version of our textbook, our strategy optimizes cost efficiency by prioritizing the use of free tiers and switching to paid instances only when absolutely necessary. This approach enables us to maintain high-quality AI-driven educational experiences for a global audience while also managing operational costs.


\begin{table*}[t]
    \centering
    \caption{Cost analysis for serverless functions using various providers and ai models.}
    \label{table:cost_analysis}
    \resizebox{\textwidth}{!}{
        \begin{tabular}{p{2.2cm} p{2.2cm} p{2.5cm} c c c c c c c c}
            \toprule
            \textbf{Provider} & \textbf{Model Host} & \textbf{Model} & \textbf{Students} & \textbf{API Calls/Day} & \textbf{Days/Week} & \textbf{Weeks} & \textbf{Total Calls} & \textbf{In Cost} & \textbf{Out Cost} & \textbf{Total Cost} \\
            \midrule
            \multicolumn{11}{c}{\textit{Estimated Monthly Costs (4 Weeks)}} \\
            \midrule
            Azure & Groq & Mixtral-8x7b & 20 & 30 & 5 & 4 & 12000 & 2.88 & 2.88 & 5.76 \\
            Azure & OpenAI & GPT-4 & 20 & 30 & 5 & 4 & 12000 & 30.00 & 120.00 & 150.00 \\
            Azure & Google & Gemini & 20 & 30 & 5 & 4 & 12000 & 0.84 & 3.60 & 4.44 \\
            AWS & Groq & Mixtral-8x7b & 20 & 30 & 5 & 4 & 12000 & 3.36 & 3.36 & 6.72 \\
            AWS & OpenAI & GPT-4 & 20 & 30 & 5 & 4 & 12000 & 36.00 & 144.00 & 180.00 \\
            AWS & Google & Gemini & 20 & 30 & 5 & 4 & 12000 & 1.00 & 4.32 & 5.32 \\
            Cloudflare & Groq & Mixtral-8x7b & 20 & 30 & 5 & 4 & 12000 & 2.40 & 2.40 & 4.80 \\
            Cloudflare & OpenAI & GPT-4 & 20 & 30 & 5 & 4 & 12000 & 25.00 & 100.00 & 125.00 \\
            Cloudflare & Google & Gemini & 20 & 30 & 5 & 4 & 12000 & 0.70 & 2.88 & 3.58 \\
            \midrule
            \multicolumn{11}{c}{\textit{Estimated Semester Costs (16 Weeks)}} \\
            \midrule
            Azure & Groq & Mixtral-8x7b & 20 & 30 & 5 & 16 & 38400 & 9.12 & 9.12 & 18.24 \\
            Azure & OpenAI & GPT-4 & 20 & 30 & 5 & 16 & 38400 & 96.00 & 384.00 & 480.00 \\
            Azure & Google & Gemini & 20 & 30 & 5 & 16 & 38400 & 2.64 & 11.52 & 14.16 \\
            AWS & Groq & Mixtral-8x7b & 20 & 30 & 5 & 16 & 38400 & 10.72 & 10.72 & 21.44 \\
            AWS & OpenAI & GPT-4 & 20 & 30 & 5 & 16 & 38400 & 115.20 & 460.80 & 576.00 \\
            AWS & Google & Gemini & 20 & 30 & 5 & 16 & 38400 & 3.20 & 13.76 & 17.00 \\
            Cloudflare & Groq & Mixtral-8x7b & 20 & 30 & 5 & 16 & 38400 & 7.68 & 7.68 & 15.36 \\
            Cloudflare & OpenAI & GPT-4 & 20 & 30 & 5 & 16 & 38400 & 80.00 & 320.00 & 400.00 \\
            Cloudflare & Google & Gemini & 20 & 30 & 5 & 16 & 38400 & 2.24 & 9.12 & 11.36 \\
            \bottomrule
        \end{tabular}
    }
\end{table*}


\subsection{Optimizations}
\label{optimizations}

The largest amount of text passed to the language models occurs during quiz generation, which requires loading information about an entire section of a chapter. However, passing an entire chapter to the language model is impractical due to token limitations and real-time requirements, which constrain the amount of text that can be processed in a single request. This challenge is significant given the length of CS249r's chapters, which average approximately 10,508 tokens, with some chapters exceeding 40,000 tokens.

\subsubsection{Token Management and Vectorization}

To keep the total tokens within a limit of \( l = 5000 \) during quiz generation, our system employs a selective text inclusion strategy. This approach keeps the input concise while preserving the context necessary for generating meaningful questions. Specifically, we vectorize only the first \( k \) sentences from each section of the chapter, where \( k \) is dynamically tuned to balance the trade-off between capturing sufficient context and adhering to the token budget.

The value of \( k \) is adjusted based on the complexity and length of the chapter. For instance, in more dense chapters like ``AI Training'' (41,434 tokens) or ``Robust AI'' (32,548 tokens), \( k \) might be smaller to ensure coverage of all critical sections. Conversely, for shorter or more introductory chapters like ``Introduction'' (14,307 tokens), a larger \( k \) value can be used to capture more detailed context.

We achieve this optimization efficiently within the browser by implementing a simplified yet fast Word Co-Occurrence Matrix to vectorize the text. This computationally lightweight solution operates entirely on the client-side, aligning with the constraints of a browser environment while ensuring robust performance. 

The details are outlined in Algorithm~\ref{algo:crypt}. This method reduces text size while retaining the most relevant information for quiz generation. It allows us to generate meaningful quizzes even for our most extensive chapters, such as those covering complex topics like privacy and security (29,728 tokens) or data engineering (26,237 tokens), without overwhelming the language model's token limit. The adaptive nature of our approach makes sure that we can maintain consistent quiz quality across our diverse range of topics, from fundamental concepts in the deep learning primer to advanced discussions in hardware acceleration and sustainable AI.

\subsubsection{Question Caching and Reuse Strategy}
\label{sec:caching}

We also implemented a question caching and reuse strategy to further optimize system performance and reduce costs. This approach is particularly effective given the substantial length of our chapters, which average approximately 7,506 tokens, with some chapters exceeding 29,000 tokens. As students engage in quizzes, our AI generates 3 to 5 questions tailored to the selected sections of the knowledge graph. These questions are saved, gradually building a comprehensive repository for each section. Once ten quizzes are accumulated per section, we initiate a balanced approach of recycling previous questions while continuing to generate new ones. If the repository reaches a predefined threshold of $n$ questions, we primarily utilize the saved questions, reducing the need for frequent AI-generated content.

This strategy significantly decreases the number of API calls to language models over time, leading to substantial cost savings and reduced computational load. The system remains responsive to user needs; if a student requests regeneration or downvotes a quiz, we discard those specific items and generate new ones, maintaining content quality and relevance. The large pool of questions enables more diverse and tailored assessments, supporting individualized learning paths even for our most extensive chapters. This dynamic and evolving repository not only enhances the learning experience, but also contributes to the long-term sustainability and scalability of the SocratiQ platform, providing valuable content for potential future updates to the textbook.

The combination of our token management strategies and this question caching system allows us to efficiently handle the varying lengths and complexities of our chapters. For instance, in more dense chapters like "Training" (29,596 tokens) or "Robust AI" (23,249 tokens), the caching system is particularly beneficial, reducing the need for frequent, token-heavy question generation. Similarly, for shorter or more introductory chapters like "Introduction" (10,219 tokens), the system can maintain a diverse question set without overreliance on the language model.

This multifaceted optimization approach ensures that SocratiQ can deliver high-quality, personalized learning experiences across our diverse range of topics, from fundamental concepts in the deep learning primer to advanced discussions in hardware acceleration and sustainable AI, while maintaining efficiency in both computational resources and costs. By intelligently balancing the use of AI-generated content with cached questions, we create a system that is both responsive to individual student needs and scalable for widespread use in various educational contexts.

\subsection{Cost Analysis}
\label{cost_analysis}

The SocratiQ AI learning companion is highly dependent on API calls to language models, which introduces an associated cost. It is important to provide an estimate of these costs to help instructors budget for deploying such a system in their classrooms.

To estimate costs, we consider the following scenario: a class of 20 students, each making 30 API calls per day, 5 days a week. Table \ref{table:cost_analysis} provides a detailed breakdown of costs between different platforms (Azure, AWS, and Cloudflare) and various language models. The table evaluates two deployment scenarios: the cost for a single month and the cost for an entire semester (16 weeks).

From the analysis, Mixtral-8x7b and Gemini emerge as the most cost-effective options in all scenarios. Semester costs for Mixtral-8x7b range from \$15.36 to \$21.44, while Gemini costs range from \$11.36 to \$17.00, depending on the provider. However, these cost savings come with trade-offs in performance. For instance, Mixtral-8x7b is a 7-billion-parameter model, which often has inferior performance compared to larger models like Gemini, estimated to have over a trillion parameters. However, GPT-4, while likely providing the highest reasoning capabilities, incurs the highest costs. Semester costs for GPT-4 typically exceed \$400.00 in most providers. This highlights a critical trade-off between cost and performance when selecting a language model for use in a classroom setting.

In particular, Cloudflare provides the lowest costs across all models, followed by Azure and AWS. It is important to emphasize that these cost estimates are based on the prices provided by the cloud providers and the AI model vendors. Actual costs may vary due to factors such as location, usage patterns, and additional expenses. Additionally, pricing and performance characteristics may change over time as new models are released and pricing evolves.



Additional cost savings are achieved through our question caching and reuse strategy, detailed in Section \ref{sec:caching}, which significantly reduces the number of API calls required over time.


\subsection{Scalability}

The open source nature of our CS249r textbook presents scalability challenges, as it potentially attracts a global audience with unpredictable usage patterns. To address this, SocratiQ is designed for efficient scalability, utilizing a ``local first'' approach that minimizes back-end demands. This design is crucial to maintain performance and cost effectiveness, regardless of whether the system is used by individual learners worldwide or in structured classroom settings.

Most of the compute-intensive tasks, such as tracking quiz progress, generating adaptive feedback, and performing quick content searches, are handled directly on the user's device. This approach significantly reduces the strain on centralized resources and ensures that the platform remains responsive, even with a large and geographically diverse user base. By pushing these operations to the client side, we can support a virtually unlimited number of simultaneous users without proportional increases in server load or costs.

For tasks requiring advanced processing, such as generating quizzes or handling complex queries, SocratiQ utilizes a serverless architecture hosted on Azure. This architecture scales automatically on demand, allowing it to accommodate spikes in user activity without requiring dedicated server infrastructure. This is particularly beneficial for an open-source textbook, where usage can fluctuate dramatically, for instance, during exam periods or when the textbook is featured in online learning communities. When usage decreases, the system seamlessly "descales," ensuring efficient resource use and lower operational costs.

\begin{algorithm}[t]
\caption{Cryptographic Hashing of PDF}
\label{algo:crypt}
\begin{algorithmic}[1]
\REQUIRE Quiz progress stats $P$
\REQUIRE Unique user ID $U$
\REQUIRE Secret key $S$ (stored on server)
\ENSURE Cryptographically hashed PDF $H$

\STATE Generate a unique UUID $u \leftarrow \text{UUID()}$
\STATE Hash the quiz progress stats and UUID: $h \leftarrow \text{Hash}(P, u)$
\STATE Combine the hashed quiz progress stats with the secret key: $k \leftarrow \text{Hash}(h, S)$
\STATE Insert the unique code $k$ at the end of the PDF content
\STATE Compute the cryptographic hash of the PDF content with the unique code: $H \leftarrow \text{Hash}(\text{PDF Content}, k)$
\STATE Store the hashed PDF content $H$ on the server, associated with the secret key $S$
\RETURN PDF with unique code $k$
\end{algorithmic}
\end{algorithm}

The combination of local device computing and serverless scalability makes it possible to provide advanced AI assistance to a wide audience, from individual self-learners across the globe to structured institutional settings, while keeping costs low and performance high. This scalability strategy not only supports the open-access philosophy of our textbook but also ensures that the quality of the learning experience remains consistent, whether the user is a single student exploring machine learning systems or part of a large cohort in a university course.

\begin{algorithm}[t]
\caption{Verification of PDF}
\begin{algorithmic}[1]
\REQUIRE PDF with unique code $k$
\REQUIRE Secret key $S$ (stored on server)
\ENSURE Verification result $V$

\STATE Extract the unique code $k$ from the end of the PDF content
\STATE Compute the cryptographic hash of the PDF content with the extracted unique code: $H' \leftarrow \text{Hash}(\text{PDF Content}, k)$
\STATE Send the hashed PDF content $H'$ and the extracted unique code $k$ to the server
\STATE On the server, compute the cryptographic hash of the PDF content with the stored secret key $S$ and the extracted unique code $k$: $H'' \leftarrow \text{Hash}(\text{PDF Content}, k)$
\IF{$H' == H''$}
\RETURN $V = \text{Verified}$
\ELSE
\RETURN $V = \text{Not Verified}$
\ENDIF
\end{algorithmic}
\end{algorithm}

In addition, this approach to scalability aligns with the evolving landscape of educational technology, where learning is increasingly happening both inside and outside traditional classroom boundaries. Designing SocratiQ to be highly scalable and adaptable allows us to support various learning scenarios. A prime example of this versatility is how SocratiQ supports both CS249r, offered locally at Harvard, and its online counterpart, the Professional Certificate in Tiny Machine Learning (TinyML) program available through HarvardX~\cite{TinyMach74:online}. While CS249r students benefit from SocratiQ in a traditional classroom setting, learners enrolled in the online TinyML program can take advantage of the same resource remotely, demonstrating how our platform seamlessly adapts to different educational contexts. This dual application showcases SocratiQ's ability to maintain its interactive and personalized learning experience across various scales, from a focused classroom environment to a massive open online course (MOOC) setting with potentially thousands of concurrent users worldwide. By supporting both local and global learning initiatives, SocratiQ shows its capacity to scale effectively while preserving the quality and personalization of the learning experience, regardless of the educational format or number of users.

\subsection{Integration with Course}

While SocratiQ can function as a standalone learning tool, its true potential is realized when integrated into the broader course structure. This integration is particularly valuable for instructors seeking to monitor and support student progress effectively. SocratiQ facilitates this by allowing students to securely share their progress with instructors through a system of cryptographically hashed PDFs.

This secure sharing mechanism serves multiple purposes. It allows instructors to gain insight into individual student engagement and performance, enabling timely interventions or support when needed. The cryptographic hash ensures the authenticity of the shared data, providing a reliable mechanism for instructors to verify the accuracy of progress reports. While enabling progress sharing, the system maintains student privacy by giving them control over what information is shared and when. Furthermore, with access to detailed progress data, instructors can adapt their teaching strategies to address common challenges or misconceptions identified through SocratiQ interactions.

The process of generating these secure PDFs is described in Algorithm~\ref{algo:crypt}, which demonstrates how hashing is employed to create tamper-evident progress reports. This approach not only ensures data integrity, but also streamlines the progress monitoring process, allowing instructors to focus on providing targeted support and enhancing the overall learning experience.

To ensure that the authenticity of the PDF file can be verified or to prevent students from altering the quiz scores offline, Algorithm 3 is used. This verification process guarantees that the PDF shared by the student has not been tampered with. The unique code $k$ acts as a digital signature, and the server verifies its authenticity by recomputing the hash and comparing it with the stored hash. This secure system provides instructors with confidence in the integrity of the data, while maintaining a seamless integration between the SocratiQ learning companion and the broader course infrastructure.

\subsection{Privacy}

Security and privacy are central to the SocratiQ platform, embodied in its ``local-first'' architecture. This design prioritizes user privacy and minimizes external data dependencies, ensuring that sensitive information remains under the user's control. 

The privacy-centric approach is the local storage of user data. All chats are securely stored in the browser's \texttt{IndexedDB}, while quiz statistics are maintained in local storage. This means that information such as progress tracking and assessment results resides on the learner's device rather than on centralized servers. By keeping data local, we significantly reduce the risk of large-scale data breaches and offer users greater control over their personal information.

When interaction with external systems is necessary, such as for section quiz requests, the data transmitted contains no personal information, maintaining user anonymity. In cases where users ask specific questions, requests do pass through our servers, but we adhere to a strict no-storage policy for these interactions.

To balance privacy with functionality and cost efficiency, we have implemented a selective centralized caching system. User-upvoted questions are saved in a centralized database, allowing us to improve the quality of generated quizzes and reduce operational costs. This approach enhances the learning experience while still safeguarding individual user data. The platform also facilitates secure progress sharing through cryptographically hashed PDFs. This feature allows learners to share their progress with instructors without compromising data integrity, striking a balance between privacy and the need for academic oversight.

By adhering to these privacy-first principles, SocratiQ ensures a strong balance between functionality, cost efficiency, and user data protection. This approach not only safeguards user information, but also fosters trust, encouraging open and honest engagement with the platform. As educational technology continues to evolve, SocratiQ's commitment to privacy positions it as a responsible and user-centric solution in the AI-assisted learning landscape.

\section{Evaluation}

We examine SocratiQ's effectiveness in supporting self-paced learning through two main approaches: an analysis of AI-generated questions and a limited case study in a sprint-style machine learning course. In the first part of our evaluation, we focus on assessing the quality of AI-generated questions. We analyze questions across various topics, using Bloom's Taxonomy to evaluate their cognitive depth and relevance to the course material. This analysis provides insights into the system's ability to generate diverse and meaningful assessments. The second part of our evaluation presents findings from a small-scale implementation of SocratiQ in a real learning environment. While limited in scope, this case study offers preliminary insights into how students interact with the system and its potential impact on engagement and learning outcomes.

\subsection{Experimental Setup}

Our experimentation was conducted in a seven-week machine learning systems course designed to maximize the benefits of self-paced learning while maintaining structured guidance. Each week featured a two-hour lecture that introduced core ML concepts, followed by student-driven exploration periods where SocratiQ provided personalized support through adaptive assessments and targeted resources. This hybrid approach blended traditional teaching methods with self-directed learning, allowing students to progress at their own pace while ensuring coverage of essential material.

\begin{figure}[t]
    \centering
    \includegraphics[width=1\linewidth]{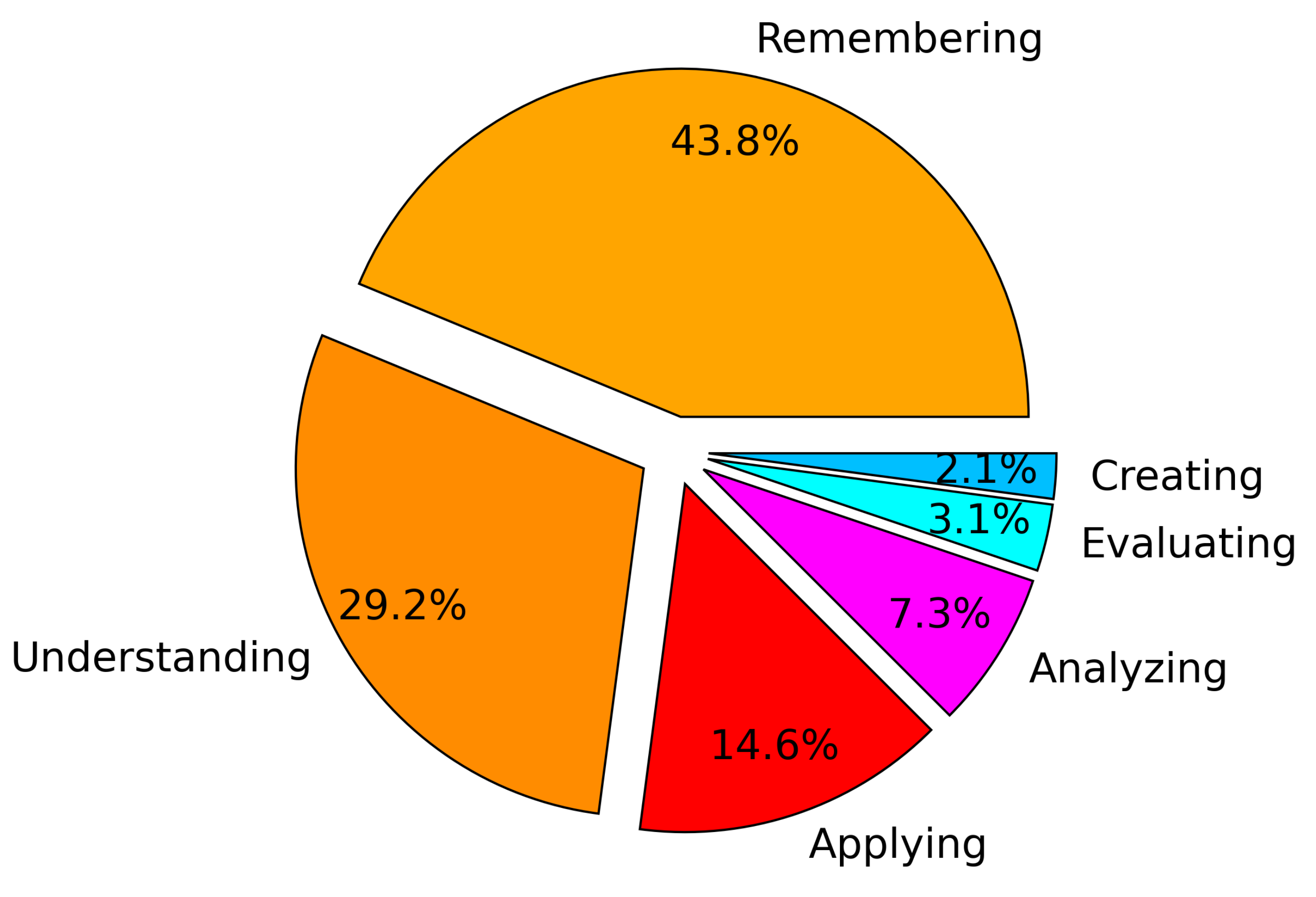}
    \caption{Distribution of generated questions by cognitive level based on Bloom's Taxonomy.}
    \label{fig:blooms-distribution}
\end{figure}

The course consisted of a small cohort of five undergraduate students with varying technical backgrounds, making self-paced learning particularly valuable. The students came from diverse academic disciplines, including computer science, engineering, and measurement technology, with most possessing intermediate programming skills. This diversity in technical proficiency was intended to showcase the ability of SocratiQ to adapt to individual learning needs while maintaining consistent educational outcomes.

SocratiQ played a central role in enabling self-paced learning through its quiz generation and resource recommendation capabilities. By aligning personalized assessments with preassigned readings, the platform allowed students to test their understanding and progress at their own pace.

\begin{table*}[t]
\centering
\caption{Examples of AI-generated questions categorized by bloom's taxonomy cognitive levels.}
\label{tab:blooms_questions}
\resizebox{\textwidth}{!}{  
\small
\begin{tabular}{l|l|p{12cm}}
\toprule
\textbf{Cognitive Level} & \textbf{Example} & \textbf{Generated Question} \\
\hline
\multirow{2}{*}{\textbf{Remembering}} 
    & Example 1 & What is the definition of a machine learning system? \\ \cline{2-3}
    & Example 2 & List three advantages of edge AI over cloud AI. \\ \hline \hline

\multirow{2}{*}{\textbf{Understanding}} 
    & Example 1 & Explain the difference between supervised and unsupervised learning. \\ \cline{2-3}
    & Example 2 & Summarize the key components of a neural network. \\ \hline \hline

\multirow{2}{*}{\textbf{Applying}} 
    & Example 1 & Identify the best algorithm for training a classification model. \\ \cline{2-3}
    & Example 2 & Use the gradient descent formula to calculate the next step for the given parameters. \\ \hline \hline

\multirow{2}{*}{\textbf{Analyzing}} 
    & Example 1 & Compare and contrast embedded AI frameworks with traditional AI frameworks. \\ \cline{2-3}
    & Example 2 & Identify potential bottlenecks in the given machine learning pipeline. \\ \hline \hline

\multirow{2}{*}{\textbf{Evaluating}} 
    & Example 1 & Assess the ethical implications of deploying facial recognition technology in public spaces. \\ \cline{2-3}
    & Example 2 & Evaluate the trade-offs between model accuracy and computational efficiency. \\ \hline \hline

\multirow{2}{*}{\textbf{Creating}} 
    & Example 1 & Design an end-to-end workflow for deploying an edge AI application. \\ \cline{2-3}
    & Example 2 & Propose a novel algorithm for handling missing data in time-series datasets. \\ \bottomrule
\end{tabular}
}
\end{table*}

\subsection{Content Quality Analysis}

To evaluate the quality and depth of AI-generated questions, we employed Bloom's Taxonomy~\cite{forehand2010bloom,krathwohl2002revision}, which categorizes educational objectives into six hierarchical levels: Remembering, Understanding, Applying, Analyzing, Evaluating, and Creating. This analysis focuses on questions generated at the beginner level, aligning with our goal of providing adaptive, personalized learning experiences as outlined in Section~\ref{sec:design}. We randomly analyzed 100 questions from the small cohort across various topics and chapters, aiming to evaluate the AI system's support for critical thinking and deep learning, identify gaps in question types, and gain insights for improving the system's question generation capabilities. This approach allows us to assess how effectively SocratiQ fulfills its role in adaptive assessment for novice learners, as described in Section~\ref{sec:design}.

Figure \ref{fig:blooms-distribution} illustrates the distribution of beginner-level questions on Bloom's cognitive levels, revealing a significant imbalance. Lower-order cognitive skills dominate, with remembering at 42\% and understanding at 28\%. In contrast, higher-order thinking skills are underrepresented: Applying (14\%), Analyzing (7\%), Evaluating (3\%), and Creating (2\%). This distribution is consistent with what can be expected for beginner-level content, focusing on building foundational knowledge and basic comprehension. The prevalence of lower-order questions at the beginner level aligns with the system's goal of providing comprehensive coverage of fundamental course material. However, it also highlights potential areas for improvement in gradually introducing higher-order thinking skills.

It is important to note that this analysis is specific to the beginner difficulty setting of SocratiQ. As shown in Table \ref{tab:chapter_difficulty_questions}, the system generates questions of varying complexity across different difficulty levels. In higher difficulty settings, we would expect to see a higher proportion of questions that target higher-order cognitive skills.

These findings have implications for SocratiQ's adaptive capabilities. Although the system provides appropriate questions for beginners, it could be improved to gradually introduce more challenging cognitive tasks even in the early stages of learning. This could help prepare learners for the more complex questions they will encounter at higher difficulty levels. Moving forward, improving SocratiQ's ability to incorporate a broader range of cognitive skills, even at the beginner level, could be beneficial. This enhancement would support a smoother transition to higher difficulty levels and potentially increase student engagement in self-paced learning environments, addressing the challenges mentioned in Section~\ref{sec:evolve}.

\subsection{AI Capability Analysis}

While our previous analysis focused on beginner-level questions, SocratiQ's AI capabilities extend beyond this single difficulty level. To provide an evaluation of the system's adaptability, we evaluated SocratiQ's performance in generating questions across various difficulty levels: Beginner, Intermediate, Advanced, and Expert, as discussed in Section~\ref{personalized_expl}. This analysis complements our earlier Bloom's Taxonomy assessment by showing how SocratiQ adapts question complexity to suit different levels of learner expertise.

Table \ref{tab:chapter_difficulty_questions} gives examples of AI-generated questions categorized by chapters and difficulty levels. This table highlights SocratiQ's ability to generate questions tailored to specific difficulty levels, ranging from introductory prompts for beginners to complex queries for advanced and expert learners. Our analysis shows several key insights about the system's performance. As the difficulty level increases from Beginner to Expert, we observe a clear progression in the complexity of questions. This progression aligns with the system's design goal of providing adaptive learning experiences. The questions show the SocratiQ's ability to adapt not only to difficulty levels but also to specific topic areas within the course. For instance, questions in the ``DNN Architectures'' chapter progress from basic definitions to intricate details of neural network operations.

\begin{table*}[h]
\centering
\caption{Examples of AI-generated questions categorized by chapters and difficulty levels.}
\label{tab:chapter_difficulty_questions}
  \resizebox{\textwidth}{!}{  
  \small
\begin{tabular}{l|l|p{11cm}}
\toprule
\textbf{Chapter} & \textbf{Difficulty Level} & \textbf{Generated Question} \\
\hline
\multirow{4}{*}{\textbf{DNN Architectures}} 
    & Beginner & What is the core operation in a Convolutional Neural Network (CNN)? \\ \cline{2-3}
    & Intermediate & What are the two outer loops in the given nested loops used for in Convolutional Neural Networks (CNNs)? \\ \cline{2-3}
    & Advanced & Memory requirements can be reduced using which type of neural network under appropriate convolutional configurations given spatially structured computations? \\ \cline{2-3}
    & Expert & Which compute pattern allows for processing a local 3x3 region of input values across all input channels through the actual convolution operation, moving across all input and output spatial positions? \\ \hline \hline

\multirow{4}{*}{\textbf{Efficient AI}} 
    & Beginner & Floating point: Known as a single-precision floating point, FP32 utilizes 32 bits to represent a number, incorporating its sign, $\_\_\_\_\_\_$ and mantissa. \\ \cline{2-3}
    & Intermediate & Why is reduced numeric precision important for hardware acceleration? \\ \cline{2-3}
    & Advanced & What are the advantages of using lower precision integer numbers during the inference phase with accelerators and GPUs optimized for those operations? \\ \cline{2-3}
    & Expert & Explain the benefits of using BF16 over regular FP16. Why is it considered an efficient alternative for deep learning applications? \\ \hline \hline

\multirow{4}{*}{\textbf{Model Optimizations}} 
    & Beginner & What is the first step in the illustrated process of network pruning? \\ \cline{2-3}
    & Intermediate & How is computational efficiency increased through structured pruning? \\ \cline{2-3}
    & Advanced & What is the main difference between iterative pruning and one-shot pruning? \\ \cline{2-3}
    & Expert & Choose a fitting statement for situations that can lead to better overall results when using one-shot pruning compared to iterative pruning. \\ \bottomrule
\end{tabular}
}
\end{table*}

Regarding cognitive skill targeting, we found that while beginner and intermediate levels tend to focus more on foundational knowledge (aligning with the lower levels of Bloom's taxonomy), advanced and expert levels provide greater opportunities for critical thinking and problem solving. This shift corresponds to higher levels of Bloom's taxonomy, such as Analyzing, Evaluating, and Creating. The language and concepts used in the questions become progressively more sophisticated and domain-specific as the difficulty level increases. For example, in the ``Efficient AI'' chapter, the beginner question asks about the basic terminology of floating-point representation, while the expert question requires learners to explain the benefits of BF16 over FP16 and its implications for deep learning applications. Similarly, in the ``Model Optimizations'' chapter, the progression moves from identifying the first step in network pruning at the beginner level to comparing one-shot and iterative pruning techniques at the expert level. This progression helps learners develop a more nuanced understanding of the subject matter. In addition, higher difficulty levels often require learners to apply their knowledge to specific scenarios or compare different concepts, fostering a deeper understanding of the material.

Despite these strengths, we also identified areas for improvement in SocratiQ's question generation capabilities. Although the system effectively differentiates between difficulty levels, there could be more gradual transitions, especially between intermediate and advanced levels. The degree of difficulty progression varies somewhat between different chapters, and standardizing this progression could enhance the overall learning experience. Furthermore, even at beginner and intermediate levels, there is potential to incorporate more questions that encourage critical thinking.

This analysis demonstrates SocratiQ's capability to generate diverse, difficulty-appropriate questions across various topics. The system's ability to adapt question complexity aligns with our goal of providing personalized learning experiences, as outlined in Section~\ref{sec:design}. However, there remains room for refinement to ensure a smoother progression of difficulty and to incorporate higher-order thinking skills more consistently across all levels. These insights will guide future improvements to SocratiQ, enhancing its ability to support learners at various stages of their educational journey.



\subsection{Student Feedback and Experiences}

To complement our quantitative analysis of SocratiQ's AI capabilities and question quality, we gathered feedback (in a systematic manner) from students across multiple topics who used the system during the sprint-style machine learning course. These qualitative data provide us with important insights into the student experience and the effectiveness of SocratiQ as a learning tool.

The \textit{ease of use and the design of the interface} of SocratiQ received mixed feedback. Although one student found it straightforward, stating:
\begin{prettiquote}
It's easy to open it and to complete a quiz.
\end{prettiquote}
Others encountered some confusion with the interface:
\begin{prettiquote}
I'm confused about why all levels of the quiz are displayed in a section, but I can only submit one specific level.
\end{prettiquote}

In terms of \textit{learning enhancement}, students found SocratiQ helpful. One student noted its effectiveness in reinforcing knowledge:
\begin{prettiquote}
It can help me deepen my memory of conceptual knowledge points.
\end{prettiquote}
Another student highlighted how the system encouraged active engagement with the material:
\begin{prettiquote}
Questions do get you to think about what you have read.
\end{prettiquote}

The \textit{AI interaction capabilities} of SocratiQ were particularly appreciated. One student shared a specific example of how the AI provided relevant information for them.
\begin{prettiquote}
When I asked him about the use of robustai, he gave me many examples.
\end{prettiquote}

Moreover, SocratiQ's ability to foster \textit{critical thinking} was emphasized by several students. One particularly insightful comment highlighted the interactive nature of the system:
\begin{prettiquote}
Yes, with the help of SocratiQ's Q\&A system, I can ask questions whenever they come up, instead of the one-way input of traditional reading. This interactive approach allows me to think more critically about the book's content.
\end{prettiquote}

When \textit{comparing SocratiQ to traditional study methods}, students found the system more engaging:
\begin{prettiquote}
SocratiQ provided an interactive experience, making studying more engaging compared to passive reading or note-taking.
\end{prettiquote}

The \textit{gamification features} of SocratiQ received mixed responses. Although some students found them motivating:
\begin{prettiquote}
This would be very effective for me because I really enjoy collecting badges.
\end{prettiquote}
Others had a more nuanced view:
\begin{prettiquote}
I liked the visual stats, not really interested in the badges, but the students might appreciate them.
\end{prettiquote}

The students also provided valuable suggestions \textit{ for improvement}. One student proposed enhancing the credibility of the system:
\begin{prettiquote}
If the related section includes not only internal answers but also external links, it might be more credible and less likely to be dismissed as an AI hallucination.
\end{prettiquote}
Another suggested adding more interactive features:
\begin{prettiquote}
Maybe add a section that can lets the students or readers discuss for each chapter, such as questions and related information.
\end{prettiquote}

These student experiences highlight both the strengths of SocratiQ and areas for potential improvement. They underscore the system's effectiveness in providing interactive, engaging learning experiences while also pointing to opportunities for enhancing user interface design, question diversity, and customization options to better cater to individual learning preferences.

The qualitative feedback largely aligns with our quantitative analysis of SocratiQ's AI capabilities. Students' appreciation of the interactive Q\&A system and immediate feedback corroborates our findings on the system's ability to generate relevant and helpful responses. However, the comments about user interface confusion and the need for more diverse content types suggest areas where further refinement could enhance the system's effectiveness in supporting self-paced learning.

\section{Discussion}

Our implementation of SocratiQ provides insight into how future educational systems might evolve, particularly as language models continue to advance at an unprecedented pace. The development of more capable models, from GPT-4 to Gemini and beyond, suggests that the capabilities shown in this paper are only the beginning of what will be possible in AI-enhanced education.

The next generation of educational AI systems will likely be shaped by several key advances. The increasing sophistication of language models will enable a more nuanced understanding of student responses and the generation of sophisticated questions, moving beyond current constraints in token management and prompt engineering. These advances could transform the way students interact with educational content, enabling more natural, conversational learning experiences. The evolution of multi-modal AI capabilities will enable seamless integration of visual, auditory, and interactive elements, particularly valuable in technical education, where understanding often requires engaging with diagrams, equations, and dynamic visualizations.

Also, we believe that cost considerations will likely shift as language model deployment becomes more efficient, with significant implications for educational access in developing nations. While our current multi-model approach using Mixtral-8x7b and Gemini helps manage expenses, future optimizations could make sophisticated AI tutoring more widely accessible. Our experiments with local model deployment and aggressive caching strategies point toward solutions for regions with limited connectivity or financial resources. Future systems could implement hybrid approaches where smaller and more efficient models run locally while maintaining access to cloud-based models when needed. The increasing availability of multilingual models further opens possibilities for delivering personalized education in local languages, addressing crucial barriers to educational access in many developing regions.

Our experience in developing SocratiQ for CS249r has already highlighted challenges in teaching complex technical content. Machine learning systems courses require students to simultaneously grasp concepts that span data engineering, model architecture, and computational infrastructure. When discussing neural network optimization, for example, students must understand the interplay between model architecture choices, hardware capabilities, and performance requirements. Future AI learning companions will need sophisticated mechanisms to dynamically adjust their depth of explanation. When a student asks about quantization, the system must determine whether to focus on mathematical principles, hardware implications, or practical implementation considerations—a challenge that becomes more complex across different institutions and student populations with varying technical backgrounds.

The integration of AI into education will likely only accelerate dramatically in the coming years. Although the architecture and implementation strategies presented here provide a foundation for future development, the field must evolve rapidly to harness new capabilities. Ultimately, the goal remains to enhance rather than replace traditional educational approaches, as discussed in Section~\ref{sec:evolve}. The next generation of AI-powered learning companions must navigate the complex intersection of technological capability and pedagogical effectiveness, ensuring that advances in AI technology translate into meaningful improvements in educational outcomes.

\section{Conclusion}

The innovative strategies we have developed and deployed within the SocratiQ platform are making significant progress in enabling AI-enhanced educational experiences. Through adaptive token management and vectorization techniques, we generate quizzes that maintain essential chapter contexts within token limits. This ensures the content's quality and relevance across a range of complex subjects while meeting real-time processing requirements. Additionally, utilizing hashing techniques to produce tamper-evident PDFs for progress reporting maintains data integrity, helping educators dedicate more time to personalized student support. The dual strategy of prioritizing free-tier access for public materials and selectively using paid instances emphasizes our commitment to providing high-quality and cost-effective education. These advancements enable the SocratiQ platform to improve learning technology, thereby promising a future of more personalized, efficient, and accessible educational content for learners around the world. As the platform continues to evolve, ongoing optimization and refinements will be critical to further enriching the student educational experience and keeping abreast of developments in AI education.

If you find value in this idea and wish to support its development, consider giving a star $\star$ to the project on GitHub. By doing so, you can help to increase its visibility and encourage further contributions from the community. Your support is essential to expand the reach and impact of AI-enhanced learning tools. Please visit the project on GitHub at \href{https://github.com/harvard-edge/cs249r_book}{\textcolor{blue} {https://github.com/harvard-edge/cs249r\_book}.}

\section*{Acknowledgments}

The authors thank the Harvard LInc Faculty Fellowship program for its support in pedagogical design and curriculum development, as well as the National Science Foundation for supporting student research through the NSF Graduate Research Fellowship under Grant No. DGE-2140743. The authors also thank the Harvard Extension School and the Harvard Data Science Initiative (HDSI) for their support in the development of SocratiQ. Furthermore, we extend our appreciation to Jeremy Ellis, Marcelo Rovai, Marco Zennaro, and the tinyML4D working group for their contributions in testing SocratiQ AI at tinyML and Edge AI workshops, particularly in developing countries, helping us advance our goal of broadening access to cutting-edge educational technologies and knowledge.



\bibliography{references}











\end{document}